\documentclass{aa}
\usepackage{graphicx}
\usepackage{color}
\usepackage[varg]{txfonts}
\graphicspath{{./Figures/}}

\begin{document}

\title{Planetesimal formation around the snow line. I\hspace{.01em}I. Dust or pebbles?}
\titlerunning{Planetesimal formation around the snow line. Dust or pebbles?}

   \author{Ryuki Hyodo
          \inst{1}
          \and
          Tristan Guillot
          \inst{2}
          \and
          Shigeru Ida
          \inst{3}
           \and
          Satoshi Okuzumi
          \inst{4}
           \and
          Andrew N. Youdin
          \inst{5}
           }
         \institute{ISAS/JAXA, Sagamihara, Kanagawa, Japan (\email{hyodo@elsi.jp})
         \and Universit\'e C\^ote d'Azur, Laboratoire J.-L.\ Lagrange, CNRS, Observatoire de la C\^ote d'Azur, F-06304 Nice, France 
         \and Earth-Life Science Institute, Tokyo Institute of Technology, Meguro-ku, Tokyo 152-8550, Japan 
         \and Department of Earth and Planetary Sciences, Tokyo Institute of Technology, Meguro-ku, Tokyo 152-8551, Japan
         \and Steward Observatory/The Lunar and Planetary Laboratory, University of Arizona, Tucson, Arizona 85721, USA. }    

\date{DRAFT:  \today}

\abstract
{Forming planetesimals is a major challenge in our current understanding of planet formation. Around the snow line, icy pebbles and silicate dust may locally pile up and form icy and rocky planetesimals via a streaming instability and/or gravitational instability. The scale heights of both pebbles and silicate dust released from sublimating pebbles are critical parameters that regulate the midplane concentrations of solids.} 
{Here, using a realistic description of the scale height of silicate dust and that of pebbles, we wish to understand disk conditions for which a local runaway pile-up of solids (silicate dust or icy pebbles) occurs inside or outside the snow line.}
{We performed 1D diffusion-advection simulations that include the back-reaction (the inertia) to radial drift and diffusion of icy pebbles and silicate dust, ice sublimation, the release of silicate dust, and their recycling through the recondensation and sticking onto pebbles outside the snow line. We used a realistic description of the scale height of silicate dust obtained from a companion paper and that of pebbles including the effects of a Kelvin-Helmholtz (KH) instability. We study the dependence of solid pile-up on distinct effective viscous parameters for turbulent diffusions in the radial and vertical directions ($\alpha_{\rm Dr}$ and $\alpha_{\rm Dz}$) and for the gas accretion to the star ($\alpha_{\rm acc}$) as well as that on the pebble-to-gas mass flux ($F_{\rm p/g}$).}
{Using both analytical and numerical approaches, we derive the sublimation width of drifting icy pebbles which is a critical parameter to characterize the pile-up of silicate dust and pebbles around the snow line. We identify a parameter space (in the $F_{\rm p/g}-\alpha_{\rm acc}-\alpha_{\rm Dz}(=\alpha_{\rm Dr})$ space) where pebbles no longer drift inward to reach the snow line due to the back-reaction that slows down the radial velocity of pebbles (we call this the "no-drift" region). We show that the pile-up of solids around the snow line occurs in a broader range of parameters for $\alpha_{\rm acc}=10^{-3}$ than for $\alpha_{\rm acc}=10^{-2}$. Above a critical $F_{\rm p/g}$ value, the runaway pile-up of silicate dust inside the snow line is favored for $\alpha_{\rm Dr}/\alpha_{\rm acc} \ll 1$, while that of pebbles outside the snow line is favored for $\alpha_{\rm Dr}/\alpha_{\rm acc} \sim 1$. Our results imply that a distinct evolutionary path in the $\alpha_{\rm acc}-\alpha_{\rm Dr}-\alpha_{\rm Dz}-F_{\rm p/g}$ space could produce a diversity of outcomes in terms of planetesimal formation around the snow line.}
 {}

\authorrunning{R. Hyodo, T. Guillot, S. Ida, S. Okuzumi, A. Youdin}
 \maketitle 

\section{Introduction} \label{sec_intro}

Forming planetesimals in protoplanetary disks is a major challenge in our current understanding of planet formation. Streaming instability (SI) \citep[e.g.,][]{You05} and gravitational instability (GI) \citep[e.g.,][]{Gol73} could be prominent candidate mechanisms to directly form planetesimals from small particles, although the detailed conditions and applicability are still a matter of debate.

\begin{figure*}[h]
	\centering
	\resizebox{\hsize}{!}{ \includegraphics{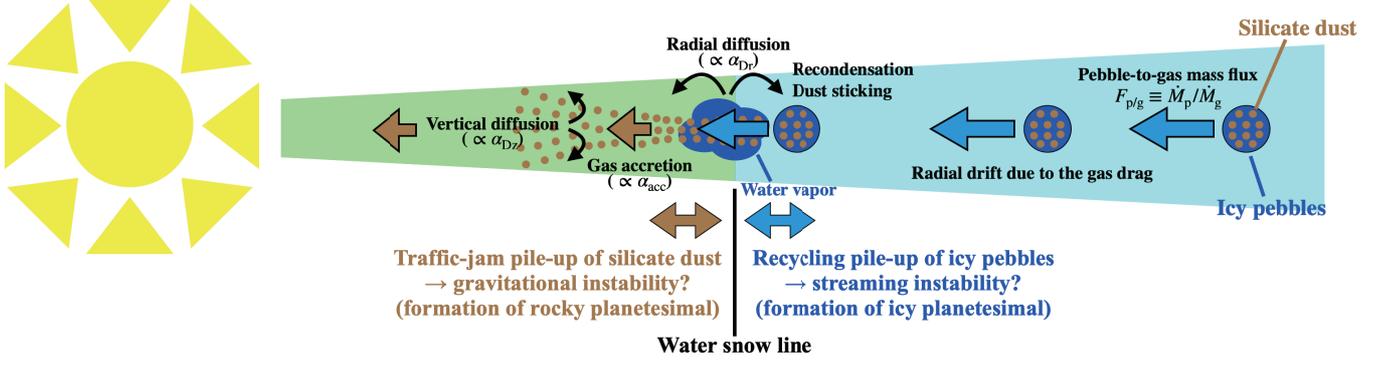} }
	\caption{Summary of solid pile-up around the snow line. Icy pebbles, which uniformly contain micron-sized silicate dust formed at the outer disk, drift inward due to the gas drag. Pebbles sublimate through the passage of the snow line (Section \ref{sec_sub}) and silicate dust is released together with the water vapor. Silicate dust is well coupled to the gas and the radial drift velocity significantly drops from that of pebbles. This causes so-called "traffic-jam" effect and the dust piles up inside the snow line. A fraction of water vapor and silicate dust diffuses from inside to outside the snow line recycles onto pebbles via the recondensation and sticking, causing a local pile-up of pebbles just outside the snow line. Midplane concentrations (i.e., midplane solid-to-gas ratio) of pebbles and silicate dust strongly depend on their scale heights. Near the snow line, the scale height of silicate dust is the minimum as the dust is released from pebbles whose scale height is much smaller than that of the gas. This causes the maximum midplane concentration of silicate dust just inside the snow line. Significant midplane concentrations of silicate dust and/or icy pebbles would cause gravitational instability and/or streaming instability, forming rocky and/or icy planetesimals, respectively (Section \ref{sec_results}).}
\label{fig_summary}
\end{figure*}

The water snow line may be a favorable location where solids selectively pile up (Fig.~\ref{fig_summary}). During the passage of icy pebbles drifting through the water snow line, icy pebbles sublimate and silicate dust is ejected\footnote{A critical collision velocity for rebound and fragmentation is conventionally expected to be $\sim 10$ times lower for silicate than for ice \citep[e.g.,][]{Blu00,Wad11}, although recent studies showed updated sticking properties of silicate \citep[e.g.,][]{Kim15,Gun18,Mus19,Ste19}.} \citep{Sai11,Mor15}. The formation of icy planetesimals by the streaming instability outside the snow line could be triggered via the local enhancement of pebble spatial density through the recondensation of diffused water vapor and sticking of silicate dust onto pebbles beyond the snow line \citep{Sch17,Dra17,Hyo19}. Inside the snow line, rocky planetesimals might be preferentially formed by the gravitational instability of piled up silicate dust because the drift velocity suddenly drops there as the size significantly decreases from sublimating pebbles \citep{Ida16, Hyo19}.

Recent 1D numerical simulations aimed to study the pile-up of solids around the water snow line \citep{Sai11,Mor15,Est16,Ida16,Sch17,Dra17,Cha19,Hyo19,Gar20}. Although these different studies considered similar settings of pebbles and silicate dust, they neglected some of the following essential physical processes to regulate the midplane solid-to-gas ratio: These are, for example, (1) gas-dust friction for solids (pebbles and/or silicate dust), that is, the back-reaction to their drift velocity and diffusive motion in the radial and vertical directions, and (2) a realistic prescription of the scale height of silicate dust as well as that of pebbles.

Back-reactions that slow down drift velocities of pebbles and silicate dust as their pile-up proceeds (hereafter, Drift-BKR) are essential for pile-up as $\Sigma_{\rm p \, (or\, d)} \propto 1/v_{\rm p \, (or\, d)}$ (a subscript of "p (or d)" describes physical parameters of either pebbles or those of silicate dust), where $\Sigma_{\rm p \, (or\, d)}$ and $v_{\rm p \, (or\, d)}$ are the surface density and the radial velocity of pebbles or silicate dust, respectively \citep{Ida16}. Back-reaction to the diffusive motion (hereafter, Diff-BKR) can trigger runaway pile-up of solids as their diffusivities are progressively weakened as pile-up proceeds \citep[][see Eqs.~(\ref{eq_Dsol_Dr}) and (\ref{eq_Dsol_Dz}) below]{Hyo19}.

The scale height of silicate dust $H_{\rm d}$ is another critical consideration on the solid pile-up. The midplane spatial density of silicate dust is $\propto 1/H_{\rm d}$ and thus the midplane solid-to-gas ratio is directly affected by $H_{\rm d}$. The efficiency of the recycling (sticking) of diffused silicate dust onto pebbles outside the snow line is  $\propto 1/H_{\rm d}$ as the number density of colliding silicate dust is $\propto 1/H_{\rm d}$. All of the previous studies of 1D calculations used a too simple prescription for $H_{\rm d}$ and their results overestimated or underestimated the pile-up of solids, depending on their assumed $H_{\rm d}$ and the disk conditions (see more details in Appendix \ref{sec_app_Hd}).

In our companion paper (\cite{Ida20}; hereafter Paper I), we performed 2D (radial-vertical) Monte Carlo simulations (a Lagrange method) of silicate dust that was released around the snow line from drifting icy pebbles under ice sublimation. Based on these Monte Carlo simulations and using semi-analytical arguments, Paper I derived a formula for the evolving scale height of the silicate dust for a given disk parameters as a function of the distance to the snow line (Eq.~(\ref{eq_Hd_Ida1}) below). The derived scale height of silicate dust is a function of the radial width where the dust is released from sublimating pebbles (i.e., sublimation width of pebbles; see Eq.~(\ref{eq_Hd_Ida1})). However, Paper I did not include the physical processes of the sublimation. Here, our 1D code is suitable for including the pressure-dependent sublimation around the snow line and we study the sublimation width of the drifting pebbles (Section \ref{sec_sub}). 

Paper I also studied conditions for the pile-up of silicate dust just inside the snow line and identified disk parameters where runaway pile-up of silicate dust occurs (i.e., rocky planetesimal formation). However, Paper I did not include recycling of water vapor and of silicate dust outside the snow line due to their diffusive motion and hence they did not consider pile-up process of pebbles outside the snow line. 

The scale height of pebbles is a function of the Stokes number (larger pebbles tend to reside in a vertically thinner disk midplane layer). However, for a sufficiently thin pebble-dominated midplane layer may interact with the upper/lower gas-dominated layer, inducing a Kelvin-Helmholtz (KH) instability \citep{Sek98,Chi08}, which was not considered before. Here, we also consider the effects of a KH instability for a more realistic picture of the scale height of pebbles and their midplane concentration (Section \ref{sec_KH}).

In reality, pile-ups of silicate dust and/or pebbles around the snow line are the consequence of combinations of ice sublimation, dust release, and their recycling onto pebbles together with their complex radial and vertical motions (Fig.~\ref{fig_summary}). In this work, we perform new 1D simulations that include both Drift-BKR and Diff-BKR \cite[the same as][]{Hyo19} and that adopt, for the first time, a realistic scale height of silicate dust (derived in Paper I) and that of pebbles (derived in Section \ref{sec_KH}). We discuss favorable conditions for runaway pile-ups of pebbles and/or silicate dust by investigating a much broader disk parameter space than previous studies.

In Section \ref{sec_methods}, we explain our numerical methods and models. In Section \ref{sec_sub}, using 1D numerical simulations and analytical arguments, we derive sublimation width of drifting icy pebbles, which is a critical parameter to regulate a local solid pile-up around the snow line. In Section \ref{sec_KH_ND}, we derive the scale height of pebbles considering the effects of a KH instability, and we also derive a critical $F_{\rm p/g}$ above which pebble drift is stopped due to the back-reaction onto the gas that slows down the radial velocity of pebbles. In Section \ref{sec_results}, we show our overall results of 1D simulations that combined with a realistic scale heights of silicate dust and pebbles. In Section \ref{sec_summary}, we summarize our paper.

\section{Numerical methods and models} \label{sec_methods}

Here, we describe numerical methods and models. In this paper, we use three distinct non-dimensional parameters to describe the gas-solid evolutions in protoplanetary disk: $\alpha_{\rm acc}$ which describes the global efficiency of angular momentum transport of gas (i.e., corresponds to the gas accretion rate) via turbulent viscosity $\nu_{\rm acc}$, $\alpha_{\rm Dr}$ which describes the radial turbulence strength (i.e., corresponds to the radial velocity dispersion), and $\alpha_{\rm Dz}$ which describes the vertical turbulence strength (i.e., corresponds to the vertical velocity dispersion) and which is used to describe the scale height of pebbles\footnote{In \cite{Hyo19}, $\alpha_{\rm tur}$ is used to describe $\alpha_{\rm Dr}$ and $\alpha_{\rm Dz}$.}. 

The classical $\alpha$-disk model \citep{Sha73} has adopted fully turbulent disks, i.e., $\alpha_{\rm acc} \simeq \alpha_{\rm Dr} \simeq \alpha_{\rm Dz}$, whereas recent theoretical developments suggest that the vertically averaged efficiency of the disk angular momentum transport, i.e., $\alpha_{\rm acc}$, could be different from local radial and vertical turbulence strengths $\alpha_{\rm Dr}$ and $\alpha_{\rm Dz}$ \citep[e.g.,][]{Arm13,Liu19}. The gas accretion can be dominated by a process other than radial turbulent diffusion such as angular momentum transport by large-scale magnetic fields \citep[e.g.,][]{Bai16}. Magnetohydrodynamic (MHD) simulations have shown that a region of ionization that is too low for the magneto-rotational instability (MRI) to operate ("dead zone"; \cite{Gam96}) might ubiquitously exist in the inner part of disk midplane and only surface layers are magnetically active \citep[$\alpha_{\rm Dr},\alpha_{\rm Dz} = 10^{-5} - 10^{-3}$ within a dead zone; e.g.,][]{Gre15,Sim15,Bai13,Bai16,Mor17}. These results indicate $\alpha_{\rm Dr}, \alpha_{\rm Dz} < \alpha_{\rm acc}$ \citep[see also][]{Zhu15,Has17,Yan18} and we study a wide parameter range where $\alpha_{\rm Dr}(=\alpha_{\rm Dz}) \leq \alpha_{\rm acc}$.

\subsection{Structure of the gas disk}
The surface density of the gas at a radial distance $r$ is expressed as a function of the gas accretion rate $\dot{M}_{\rm g}$ and the gas radial velocity $v_{\rm g}$ (i.e., the classical $\alpha$-accretion disk model; \cite{Sha73,Lyn74}) as
\begin{equation}
\label{eq_sigma}
	\Sigma_{\rm g}=\frac{\dot{M}_{\rm g}}{2\pi r v_{\rm g}}.
\end{equation}
The isothermal sound speed of the gas is written as
\begin{equation}
\label{eq_sound}
	c_{\rm s} = \sqrt{ \frac{k_{\rm B} T}{\mu_{\rm g}m_{\rm proton}}} ,
\end{equation}
where $k_{\rm B}$ is the Boltzmann constant, $\mu_{\rm g}$ is the mean molecular weight of the gas ($\mu_{\rm g}=2.34$), $m_{\rm proton}$ is the proton mass, and $T$ is the temperature of the disk gas. In this paper, the temperature profile of the disk is fixed as 
\begin{equation}
\label{eq_temp}
	T(r) = T^{*} \left( \frac{r}{3.0 \, \rm au} \right)^{-\beta},
\end{equation}
where we use $T^{*}=150$ K and $\beta=1/2$. Then, the scale height of the gas is given as
\begin{equation}
	H_{\rm g} = \frac{c_{\rm s}}{\Omega_{\rm K}} \simeq 0.033 \, \rm au \times  \left( \frac{r}{1.0 \, \rm au} \right)^{5/4},
\end{equation}
where $\Omega_{\rm K}$ is the Keplerian orbital frequency.
Disk gas pressure at the midplane $P_{\rm g}$ is given as
\begin{equation}
	P_{\rm g} = \frac{\Sigma_{\rm g}}{\sqrt{2\pi}H_{\rm g}} \frac{k_{\rm B}T}{\mu_{\rm g} m_{\rm proton}} .
\end{equation}
The gas rotates at sub-Keplerian speed. The degree of deviation of the gas rotation frequency from that of Keplerian $\eta$ is given by
\begin{equation}
	\eta \equiv \frac{\Omega_{\rm K} - \Omega}{\Omega_{\rm K}} = -\frac{1}{2} \frac{\partial \ln P_{\rm g}}{\partial \ln r} \left( \frac{H_{\rm g}}{r} \right)^2 = C_{\rm \eta} \left( \frac{H_{\rm g}}{r} \right)^2 ,
\end{equation}
where $\Omega$ is the orbital frequency of the gas and we define 
\begin{equation}
	C_{\rm \eta} \equiv -\frac{1}{2} \frac{\partial \ln P_{\rm g}}{\partial \ln r} ,
\end{equation}
which depends on the gas structure. In this paper, $C_{\rm \eta} = 11/8$ for $\Sigma_{\rm g} \propto r^{-1}$ and $T \propto r^{-1/2}$.

\subsection{Radial drifts of gas, pebbles, and silicate dust}
The gas radial velocity including the effects of gas-solid friction is given as \citep{Ida16,Sch17}
\begin{equation}
\label{eq_vg_BK}
	v_{\rm g,BK} = \frac{\Lambda}{1 + \Lambda^2 \tau_{\rm s}^2} \left( 2  Z \Lambda \tau_{\rm s}\eta v_{\rm K} + \left( 1+\Lambda \tau_{\rm s}^2 \right) v_{\rm g,\nu} \right) ,
\end{equation}
where $Z$ is the midplane solid-to-gas ratio of pebbles or silicate dust, $\Lambda$ characterizes the strength of the back-reaction due to pile-up of either pebbles or silicate dust, $\tau_{\rm s}$ is the Stokes number of pebbles or silicate dust, and $v_{\rm g,\nu} $ is an unperturbed disk gas accretion velocity. $Z$ is defined as
\begin{equation}
\label{eq_Z}
	Z \equiv \frac{\rho_{\rm p \, (or\, d)}}{\rho_{\rm g}} ,
\end{equation}
where $\rho_{\rm g}=\Sigma_{\rm g}/\sqrt{2\pi}H_{\rm g}$ and $\rho_{\rm p \, (or\, d)}=\Sigma_{\rm p \, (or\, d)}/\sqrt{2\pi} H_{\rm p \, (or\, d)}$ are the midplane spatial density of the gas and that of pebbles (or silicate dust), respectively. $H_{\rm p}$ and $H_{\rm d}$ are the scale heights of pebbles and silicate dust, respectively. $\Lambda$ is defined as
\begin{equation}
\label{eq_lambda}
	\Lambda \equiv \frac{\rho_{\rm g}}{\rho_{\rm g} + \rho_{\rm p \, (or\, d)}} = \frac{1}{1+Z}.
\end{equation}
Using $H_{\rm p}$ or $H_{\rm d}$, $\Lambda$ is rewritten as 
\begin{align}
\label{eq_lambda_inv}
	\Lambda^{-1} &=1 + \frac{\rho_{\rm p \, (or\, d)}}{\rho_{\rm g}}\\
	&\simeq 1 + \left(  \frac{H_{\rm g}}{H_{\rm p \, (or\, d)}} \right) \left( \frac{ \Sigma_{\rm p \, (or\, d)} }{\Sigma_{\rm g}} \right) = 1 + h_{\rm p/g \, (or \, d/g)}^{-1} Z_{\Sigma} ,
\end{align}
where $h_{\rm p/g \, (or \, d/g)} \equiv H_{\rm p \, (or\, d)}/H_{\rm g}$. $Z_{\Sigma} \equiv \Sigma_{\rm p \, (or\, d)}/\Sigma_{\rm g}$ is the vertically averaged metallicities of pebbles or silicate dust. Using the dimensionless effective viscous parameter $\alpha_{\rm acc}$ and the effective viscosity $\nu_{\rm acc}=\alpha_{\rm acc}c_{\rm s}^2 \Omega_{\rm K}^{-1}$, $v_{\rm g,\nu} $ is given as 
\begin{align}
\label{eq_vg_nu}
	v_{\rm g,\nu} &= -\frac{3\nu_{\rm acc}}{r} \frac{\partial \ln (r^{1/2} \nu_{\rm acc} \Sigma_{\rm g})}{\partial \ln r} =  -\frac{3\nu_{\rm acc}}{2r} \left( 1+ 2Q \right) ,
\end{align}
where $Q=\frac{\partial \ln (\nu_{\rm acc} \Sigma_{\rm g})}{\partial \ln r}$ \citep{Des17}. For a steady-state disk, the gas accretion velocity is $v_{\rm g,\nu}^{*} = -\frac{3\nu_{\rm acc}}{2r}$ and is rewritten as  
\begin{align}
	v_{\rm g,\nu}^{*} &= -\frac{3\nu_{\rm acc}}{2r} = -\frac{3\alpha_{\rm acc} H_{\rm g}^{2} \Omega_{\rm K}}{2 r} = -\frac{3\alpha_{\rm acc}}{2} \left( \frac{H_{\rm g}}{r} \right)^{2} v_{\rm K}\\ &\simeq -\frac{3}{2} \alpha_{\rm acc} \eta  v_{\rm K} \left( -\frac{1}{2} \frac{\partial \ln P_{\rm g}}{\partial \ln r} \right)^{-1} = -\frac{3}{2} \alpha_{\rm acc} \eta  v_{\rm K} C_{\rm \eta}^{-1} ,
\end{align}
where $v_{\rm K}$ is the Keplerian velocity.

In this work, we adopt a vertical two-layer model to describe the radial velocity of the gas: a pebble/dust-rich midplane layer with scale height $H_{\rm p \, (or\, d)}$ and a pebble/dust-poor upper layer. The vertically averaged radial velocity of the gas $v_{\rm g}$ is given as
\begin{align}
\label{eq_vg}
	v_{\rm g,p \, (or\, d)} = \frac{v_{\rm g,BK}H_{\rm p \, (or\, d)} + \left(H_{\rm g} - H_{\rm p \, (or\, d)} \right) v_{\rm g,\nu} }{H_{\rm g}}.
\end{align}
We note that the above simple two-layer model well reproduces the results of \cite{Kan17} where the approximation of the Gaussian distribution is used.

The radial velocities of pebbles and silicate dust including the effects of gas-solid friction $-$ drift back-reaction (Drift-BKR) $-$ are given as \citep{Ida16,Sch17,Hyo19}
\begin{align}
\label{eq_vp}
		v_{\rm p} &= - \frac{\Lambda}{1+\Lambda^{2}\tau_{\rm s,p}^{2}} \left( 2\tau_{\rm s,p}\Lambda \eta v_{\rm K} - v_{\rm g,\nu} \right)\\
\label{eq_vd}	
		v_{\rm d} &= - \frac{\Lambda}{1+\Lambda^{2}\tau_{\rm s,d}^{2}} \left( 2\tau_{\rm s,d}\Lambda \eta v_{\rm K} - v_{\rm g,\nu} \right) ,
\end{align}
where $\tau_{\rm s,p}$ and $\tau_{\rm s,d}$ are the Stokes number of pebbles and silicate dust, respectively. 

\subsection{Diffusion of gas and solids}
 The radial and vertical diffusivities of gas and vapor ($D_{\rm Dr,g}$ and $D_{\rm Dz,g}$) are given as
\begin{align}
\label{eq_D_gas}
	&D_{\rm Dr,g} =  \nu_{\rm Dr} = \alpha_{\rm Dr}c^{2}_{\rm s} \Omega_{\rm K}^{-1}\\
	&D_{\rm Dz,g} =  \nu_{\rm Dz}  = \alpha_{\rm Dz}c^{2}_{\rm s} \Omega_{\rm K}^{-1} ,
\end{align}
where $\nu_{\rm Dr}$ and $\nu_{\rm Dz}$ are the turbulent viscosities that regulate radial and vertical diffusions in association with dimensionless turbulence parameters $\alpha_{\rm Dr}$ and $\alpha_{\rm Dz}$, respectively.

The nature of solid diffusion in the gas is partial coupling with the gas eddies \citep{You07}. \cite{Hyo19} considered that the diffusivity of pebbles (and that of silicate dust) is reduced owing to the gas-solid friction as pile-up of solids proceeds $-$ diffusion back-reaction (Diff-BKR) $-$ and radial and vertical diffusivities of pebbles and silicate dust are given as
\begin{equation}
\label{eq_Dsol_Dr}
	D_{\rm Dr,p \, (or\, d)} = D_{\rm Dr,g} \times \frac{ \Lambda^{K} }{1+{\rm \tau_{\rm s,p \, (or\, d)}}^2}
\end{equation}
\begin{equation}
\label{eq_Dsol_Dz}
	D_{\rm Dz,p \, (or\, d)} = D_{\rm Dz,g} \times \frac{ \Lambda^{K} }{1+{\rm \tau_{\rm s,p \, (or\, d)}}^2} ,
\end{equation}
where $K$ is the coefficient. $K=1$ could be applied because the diffusivity is $D_{\rm g} \propto c^{2}_{\rm s} \propto 1/\rho_{\rm g}$ under a constant $P$ assumption. Using the effective gas density, $\rho_{\rm g,eff}=\rho_{\rm g}+\rho_{\rm p \, (or\, d)}$, the effective sound velocity of a mixture of gas and small particles is given as
\begin{equation}
	c^{2}_{\rm s,eff} = c^{2}_{\rm s,K=0} \left( \frac{\rho_{\rm g}}{\rho_{\rm g}+\rho_{\rm p \, (or\, d)}} \right) = c^{2}_{\rm s,K=0} \Lambda ,
\end{equation} 
where $c_{\rm s,K=0}$ is the sound velocity when $K=0$ (Eq.~(\ref{eq_sound})). However, from an energy dissipation argument, the lower limit is $K = 1/3$. We note that the result of solid pile-up around the snow line does not significantly depend on the choice of $K$ as long as the back reaction to diffusion is included (i.e., $K > 0$). In this work, we use $K=1$ as a representative case.

\subsection{Equations of radial transport of gas, vapor, and solids}
In this paper, we solve the radial motions of gas, pebbles, silicate dust, and water vapor. Pebbles are modeled as a homogenous mixture of water ice and micron-sized silicate dust \citep[see also][]{Ida16,Sch17,Hyo19}. Due to the sublimation of icy pebbles near the snow line, water vapor and silicate dust are produced.

The governing equations of the surface density of the gas $\Sigma_{\rm g}$, that of pebbles $\Sigma_{\rm p}$, that of silicate dust $\Sigma_{\rm d}$, that of water vapor $\Sigma_{\rm vap}$, and the number density of pebbles $N_{\rm p}$ are given as \citep{Des17,Hyo19};
\begin{align}
\label{eq_sigma_ice_peb}
	& \frac{\partial \Sigma_{\rm g}}{\partial t} = -\frac{1}{r} \frac{\partial}{\partial r} \left( r \Sigma_{\rm g} v_{\rm g} \right)\\
\label{eq_sigma_ice_peb}
	& \frac{\partial \Sigma_{\rm p}}{\partial t} + \frac{1}{r}\frac{\partial}{\partial r} \left( r \Sigma_{\rm p} v_{\rm p} - r D_{\rm Dr,p}\Sigma_{\rm g} \frac{\partial}{\partial r} \left( \frac{\Sigma_{\rm p}}{\Sigma_{\rm g} } \right) \right) =\dot{\Sigma}_{\rm p}\\
	& \frac{\partial \Sigma_{\rm d}}{\partial t} +  \frac{1}{r}\frac{\partial}{\partial r}  \left( r \Sigma_{\rm d} v_{\rm d} - r D_{\rm Dr,d}\Sigma_{\rm g}  \frac{\partial}{\partial r}  \left( \frac{\Sigma_{\rm d}}{\Sigma_{\rm g} } \right) \right) =\dot{\Sigma}_{\rm d}\\
\label{eq_sigma_vap}
	& \frac{\partial \Sigma_{\rm vap}}{\partial t} +  \frac{1}{r}\frac{\partial}{\partial r}  \left( r \Sigma_{\rm vap} v_{\rm g,\nu} - r D_{\rm Dr,g}\Sigma_{\rm g}  \frac{\partial}{\partial r}  \left( \frac{\Sigma_{\rm vap}}{\Sigma_{\rm g} } \right) \right) =\dot{\Sigma}_{\rm vap}\\
\label{eq_Np}
	& \frac{\partial N_{\rm p}}{\partial t} +  \frac{1}{r}\frac{\partial}{\partial r}  \left( r N_{\rm p} v_{\rm p} - r D_{\rm Dr,p}N_{\rm g}  \frac{\partial}{\partial r}  \left( \frac{N_{\rm p}}{N_{\rm g} } \right) \right) = 0.
\end{align}
The right-hand sides of Eqs.~(\ref{eq_sigma_ice_peb})-(\ref{eq_sigma_vap}) are due to sublimation of icy pebbles, the recondensation of water vapor, and sticking of silicate dust onto pebbles (see \cite{Hyo19}). The evolution of $N_{\rm p}$ needs to be solved because the size of pebbles changes during sublimation and condensation, and the mass of pebbles is given as $m_{\rm p} = \Sigma_{\rm p}/N_{\rm p}$ under the single-size approximation.

The decrease rate of the pebble mass $m_{\rm p}$ is given by \citep{Lic91,Ros19}
\begin{equation}
	\frac{dm_{\rm p}}{dt} = 4\pi r_{\rm p}^2 v_{\rm th} (\rho_{\rm vap} - \rho_{\rm sat}) , 
\label{eq_subl0}
\end{equation}
where $r_{\rm p}$ is the particle physical radius, $\rho_{\rm vap}$ and $\rho_{\rm sat}$ are vapor and saturation spatial densities, and $v_{\rm th}$ is the averaged normal component of the velocity passing through the particle surface. The averaged normal velocity of only outgoing particles with Maxwell distribution is 
\begin{align}
	v_{\rm th} = \frac{c_{\rm s}}{\sqrt{2\pi}} = \frac{1}{\sqrt{2\pi}} \left(\frac{k_{\rm B} T}{\mu m_{\rm proton}}\right)^{1/2} , 
\end{align}
where $\mu$ is the mean molecular weight. Another definition of the thermal velocity is the averaged value of the magnitude of three-dimensional velocity, which is given by 
\begin{align}
	v_{\rm th,3D} = \sqrt{\frac{8}{\pi}} c_{\rm s} .
\end{align}
Because this definition of $v_{\rm th}$ does not include the effect of oblique ejection/collisions, the cross section must be considered rather than the surface density, that is, 
\begin{equation}
	\frac{dm_{\rm p}}{dt} = \pi r_{\rm p}^2 v_{\rm th,3D} (\rho_{\rm vap} - \rho_{\rm sat}).
\label{eq_subl1}
\end{equation}
Both Eqs.~(\ref{eq_subl0}) and (\ref{eq_subl1}) give the same $dm_{\rm p}/dt$. We note that \cite{Sch17} and \cite{Hyo19} used $v_{\rm th,3D}$ for Eq.~(\ref{eq_subl0}), which caused overestimation of the sublimation and recombination rates by a factor of 4.

\subsection{Scale heights of pebbles and silicate dust}
The scale heights of pebbles and silicate dust describe the degree of concentration of solids in the disk midplane ($\rho_{\rm p \, (or\, d)} \propto H^{-1}_{\rm p \, (or\, d)}$).

In the steady-state, the scale height of pebbles is regulated by the vertical turbulent stirring $H_{\rm p,tur}$ \citep{Dub95,You07,Oku12} as\footnote{Previous studies \citep{Sch17,Dra17,Hyo19} did not consider the effects of Diff-BKR on the scale height of pebbles (i.e., $K=0$ is assumed for Eq.~(\ref{eq_Hp_tur})).}
\begin{equation}
\label{eq_Hp_tur}
	H_{\rm p,tur} = \left( 1 + \frac{\tau_{\rm s,p}}{\alpha_{\rm Dz} \left( 1+Z_{\rm p} \right )^{-K}} \right)^{-1/2} H_{\rm g}.
\end{equation}
However, for small $\alpha_{\rm Dz}$ (i.e., for a small $H_{\rm p,tur}$), a vertical shear Kelvin-Helmholtz (KH) instability would prevent the scale height of pebbles from being a smaller value. The scale height of pebbles regulated by a KH instability $H_{\rm p, KH}$ is given (see the derivation in Section \ref{sec_KH}) as 
\begin{equation}
	H_{\rm p,KH} \simeq Ri^{1/2} \frac{Z^{1/2}}{ \left( 1+Z \right)^{3/2} } C_{\rm \eta} \left( \frac{H_{\rm g}}{r} \right) H_{\rm g},
\label{eq_Hp_KH}
\end{equation}
where $Ri=0.5$ is used here as a critical value for a KH instability to operate.
Thus, the scale height of pebbles $H_{\rm p}$ is given as
\begin{equation}
\label{eq_Hp}
	H_{\rm p} = \max \left\{ H_{\rm p,tur}, H_{\rm p,KH} \right\}.	
\end{equation}

Using a Monte Carlo simulations, Paper I derived the scale height of silicate dust $H_{\rm d}$ as a function of the scaled distance to the snow line $\Delta \tilde{x}_{\rm snow} = (r - r_{\rm snow})/H_{\rm g}$ and the scaled sublimation width $\Delta \tilde{x}_{\mathrm{subl}} = \Delta x_{\mathrm{subl}}/H_{\rm g}$ (Section \ref{sec_sub}) as 
\begin{equation}
\label{eq_Hd_Ida1}
	H_{\rm d} \simeq \left(  h_{\rm d/g,0}^{-1} + h_{\rm d/g,*}^{-1} \right)^{-1} H_{\rm g} , 
\end{equation}
where
\begin{equation}
\label{eq_Hd_Ida2}
	h_{\rm d/g,0} = \left( 1 + \frac{\tau_{\rm s,d}}{\alpha_{\rm Dz}} \right)^{-1/2} , 
\end{equation}
and
\begin{align}
	h_{\rm d/g,*} \simeq & \left( h_{\rm p/g,0}^2 + \frac{2}{3} \frac{\alpha_{\mathrm{Dz}}}{\alpha_{\mathrm{acc}}} \frac{\Delta \tilde{x} }{H_{\rm g} / r}\right)^{1/2} \nonumber \\
	 & \times \left(1+\frac{2}{3} \frac{\alpha_{\rm Dr}/\alpha_{\mathrm{acc}}}{1+\left( C_{\rm r.diff} \alpha_{\rm Dr}/\alpha_{\rm acc} \right)^2} \frac{1}{\left(H_{\rm g} / r\right) \left( \Delta \tilde{x} + \epsilon \right) }\right),
\label{eq_Hd_Ida3}
\end{align}
where $C_{\rm r.diff}=10$, $\Delta \tilde{x} = \max \{ \Delta \tilde{x}_{\mathrm{subl}}, \Delta \tilde{x}_{\rm snow} \}$, and $\epsilon=0.01$ is a softening parameter that prevents unphysical peak for very small $\Delta \tilde{x}_{\rm subl}$ at the snow line ($\Delta \tilde{x}_{\rm snow} = 0$), respectively (Paper I). The scale height of pebbles is given at the snow line (with $\tau_{\rm s,p}=0.1$) as 
\begin{equation}
	h_{\rm p/g,0} = \max \left\{ \frac{H_{\rm p,tur} \mid_{K=0}}{H_{\rm g}}, \frac{H_{\rm p,KH} \mid_{Z=0.5}}{H_{\rm g}} \right\} .
\end{equation}
Previous works adopted different simplifying prescriptions for $H_{\rm d}$ \citep[e.g.,][]{Ida16,Sch17,Hyo19}, leading to very different results for the resultant pile-up of silicate dust and pebbles (Appendix~\ref{sec_app_Hd}). Equations (\ref{eq_Hd_Ida1})--(\ref{eq_Hd_Ida3}) provide a more realistic treatment based on the 2D calculation from Paper I. 

\subsection{Numerical settings} \label{sec_settings}
Following Paper I, the ratio of the mass flux of pebbles $\dot{M}_{\rm p}$ to that of the gas $\dot{M}_{\rm g}$ is a parameter and we define it as 
\begin{equation}
	F_{\rm p/g} \equiv \frac{\dot{M}_{\rm p}}{\dot{M}_{\rm g}} .
\end{equation}

We set $\dot{M}_{\rm g}=10^{-8} M_{\odot}$/year. The inner and outer boundaries of our 1D simulations are set to $r_{\rm in}=0.1$ au and $r_{\rm out}=5.0$ au, respectively. We set the initial constant size of pebbles (radius of $r_{\rm p}$) with $\tau_{\rm s,p}(r)=0.1$ at $r=4$ au before sublimation takes place \citep[][Appendix \ref{sec_appendix_size}]{Oku12,Ida16}. The mass fraction of silicate in icy pebbles is set to $f_{\rm d/p} =0.5$ and we modeled that silicate dust uniformly embedded in icy pebbles is micron-sized small grains \citep{Sch17,Hyo19,Ida20}. The surface density of pebbles at the outer boundary is set by considering Drift-BKR and Diff-BKR (see Section \ref{sec_ND} and Eq.~(\ref{eq_peb_pileup})), which is fixed during the simulations. We consider sublimation/condensation of ice as well as release/recycling of silicate dust around the snow line \citep[][]{Hyo19}. 1D simulations are initially run without back-reactions to reach a steady-state ($\sim 10^5$ years) and then we turn on the back-reactions (Drift-BKR and Diff-BKR with $K=1$) to see the further evolution (up to $5 \times 10^5$ years).   

\begin{figure*}[h]
\resizebox{\hsize}{!}{ \includegraphics{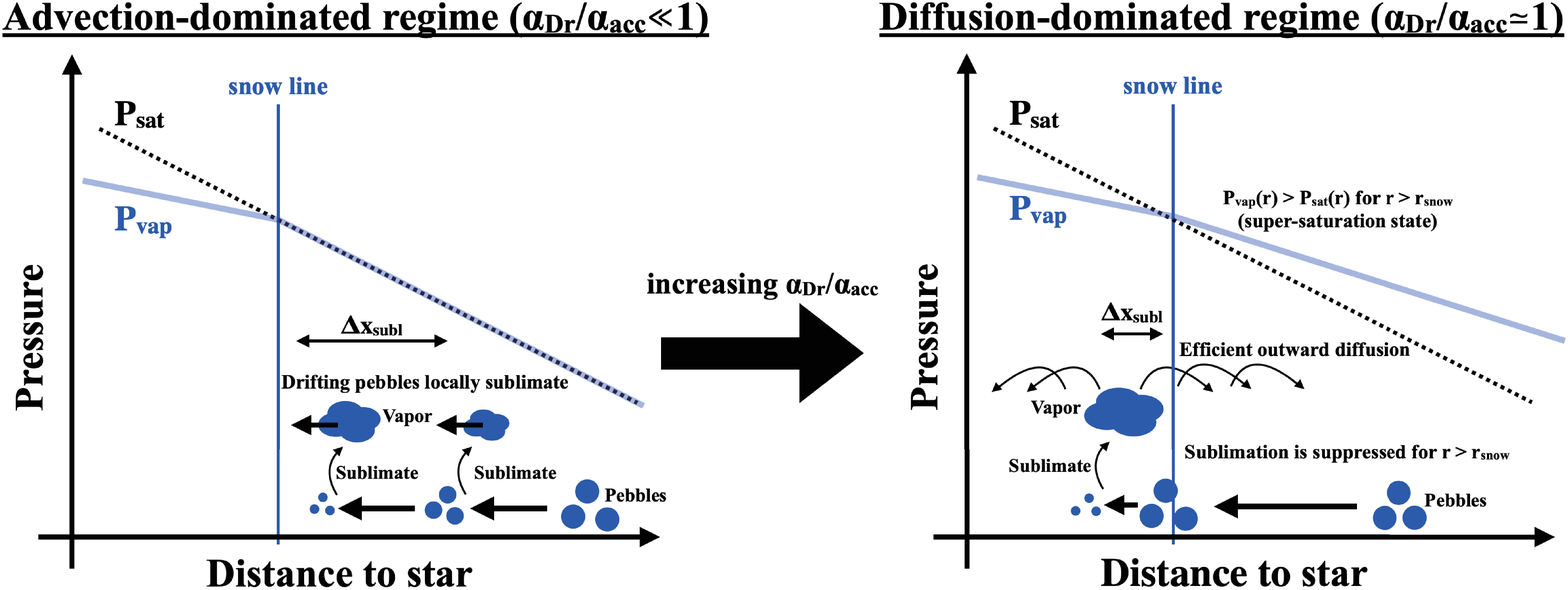}}
\caption{Schematic illustration of the sublimation of drifting icy pebbles. In the case of the advection-dominated regime (left; Section \ref{sec_advection_dominated}), gas accretion onto the central star dominates the radial transport of vapor. The loss of vapor due to the inward advection of the vapor then balances the vapor supply from sublimating icy pebbles. Thus, drifting pebbles sublimate outside the snow line (i.e., $r > r_{\rm snow}$). In the case of the diffusion-dominated regime (right; Section \ref{sec_diffusion_dominated}), water vapor produced inside the snow line efficiently diffuses outside the snow line, leading to a super-saturation state ($P_{\rm vap}(r) > P_{\rm sat}(r)$). The sublimation of pebbles can therefore take place only inside the snow line (i.e., $r < r_{\rm snow}$).}
\label{fig_sub_schematic}
\end{figure*}
 
\section{Sublimation of drifting icy pebbles} \label{sec_sub}

A description of $H_{\rm d}$ is critical to understand the resultant pile-up of silicate dust inside the snow line. Also, $H_{\rm d}$ describes the efficiency of recycling of silicate dust onto icy pebbles and pile-up of pebbles just outside the snow line because its efficiency is $\propto H_{\rm d}^{-1}$ \citep{Hyo19} (see Appendix \ref{sec_app_Hd} for a detailed comparison of different models of $H_{\rm d}$). Paper I derived a semi-analytical expression of $H_{\rm d}$ that depends on the sublimation width $\Delta x_{\rm subl}$ (see Eq.~(\ref{eq_Hd_Ida1})). Although the global distribution of water within protoplanetary disks has been a topic of research \citep[e.g.,][]{Cie06}, the detailed study of the radial width where drifting pebbles sublimate around the snow line (i.e., sublimation width) has not been conducted yet.

Below, we discuss the radial width of sublimation of drifting icy pebbles $\Delta x_{\rm subl}$ in a protoplanetary disk considering the local temperature-pressure environment. Here, we consider the cases of pure icy pebbles (i.e., $f_{\rm d/p}=0$ and $\tau_{\rm s,p}=0.1$ at 4 au). We neglect recycling processes of the recondensation of water vapor as well as sticking of silicate dust onto pebbles outside the snow line via their diffusive processes, so that we can purely focus on the sublimation process. 

In the 1D simulations, diffusion radially mixes different-sized sublimating pebbles near the snow line, which comes from solving $N_{\rm d}$ evolution (i.e., the effective size of pebbles at a given radial distance is calculated). In this section, to suppress these mixing effects and to see the size evolution of a single drifting pebble, we set $\alpha_{\rm Dr}/\alpha_{\rm acc}=10^{-3}$ for pebbles and silicate dust, while we change $\alpha_{\rm Dr}/\alpha_{\rm acc}$ for the evolution of vapor. The arguments in this section are applicable to the 1:1 rock-ice-mixed pebbles (see Section \ref{sec_settings}) because the size and density of pebbles are expected to only weakly change with slightly different compositions.

The sublimation of drifting icy pebbles takes place as long as $P_{\rm vap}(r) < P_{\rm sat}(r)$ where $P_{\rm vap}(r)$ is the water vapor pressure and $P_{\rm sat}$ is the saturation vapor pressure (Fig.~\ref{fig_sub_schematic}). $P_{\rm vap} > P_{\rm sat}$ indicates a super-saturation state. The saturation vapor pressure of water is given as
\begin{equation}
	P_{\rm sat}(r) = P_{\rm 0} \exp \left( - \frac{T_{0}}{T(r)} \right) ,
\end{equation}
where $P_{\rm 0} = 1.14 \times 10^{13}$ g cm$^{-1}$ s$^{-2}$ and $T_{\rm 0} = 6062$ K \citep{Lic91}, which can be rewritten as
\begin{equation}
	\log_{10} \left( \frac{P_{\rm sat}(r)}{\rm dyn \, cm^{-2}} \right) \simeq 13.06 - \frac{2632 \, \rm K}{T(r)}.
\end{equation}

\begin{figure}[h]
\resizebox{\hsize}{!}{ \includegraphics{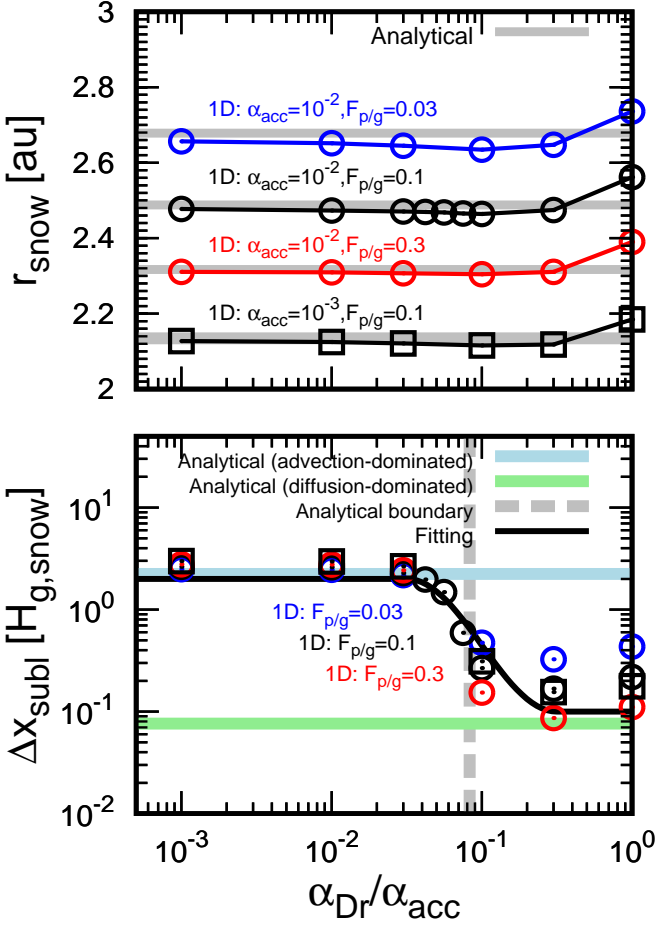}}
\caption{Radial location of the snow line (top panel) and sublimation width of drifting pebbles $\Delta x_{\rm subl}$ (bottom panel) as a function of $\alpha_{\rm Dr}/\alpha_{\rm acc}$. Here, $f_{\rm d/p}=0$ (i.e., pure icy pebbles). Circles ($\alpha_{\rm acc}=10^{-2}$) and squares ($\alpha_{\rm acc}=10^{-3}$) represent the results of 1D simulations. Blue, black and red colors indicate those for $F_{\rm p/g}=0.03$, $0.1$ and $0.3$, respectively. Analytically derived locations of the snow lines (solving Eq.~(\ref{eq_Ps_Pvap})) are shown by the gray lines (top panel). The light-blue line in the bottom panel shows the analytical $\Delta x_{\rm subl}$ in the case of the advection-dominated regime (Eq.~(\ref{eq_x_subl_adv}) with $r_{\rm snow}=2.4$ au and $\beta=0.5$). The light-green line in the bottom panel shows the analytical $\Delta x_{\rm subl}$ in the case of the diffusion-dominated regime (Eq.~(\ref{eq_Dr3}) with $\tau_{\rm s,p}=0.06$, $r_{\rm p}=2.3$ cm, $r_{\rm snow}=2.4$ au). The gray dashed line in the bottom panel shows the analytically derived boundary between the advection- and diffusion-dominated regimes (Eq.~(\ref{eq_boundary}) with $r = 2.4$ au and $\beta = 0.5$). A fitting function is shown by a black line (Eq.~(\ref{eq_fitting})).}
\label{fig_sub_width}
\end{figure}

As shown in Fig.~\ref{fig_sub_schematic}, the sublimation width of drifting pebbles can be divided into two regimes depending on $\alpha_{\rm Dr}/\alpha_{\rm acc}$: the advection-dominated regime (Section \ref{sec_advection_dominated}; $\Delta x_{\rm subl} \simeq 2 H_{\rm g}$ for $\alpha_{\rm Dr} \ll \alpha_{\rm acc}$) and the diffusion-dominated regime (Section \ref{sec_diffusion_dominated}; $\Delta x_{\rm subl} \simeq 0.2 H_{\rm g}$ for $\alpha_{\rm Dr} \sim \alpha_{\rm acc}$). Figure \ref{fig_sub_width} shows the results of 1D simulations as a function of $\alpha_{\rm Dr}/\alpha_{\rm acc}$ (the top panel being a radial distance of the snow line and the bottom panel being the sublimation width). Here, the snow line $r_{\rm snow}$ is defined by the innermost radial distance where $P_{\rm vap}(r)=P_{\rm sat}(r)$. The sublimation width $\Delta x_{\rm subl}$ is defined by the radial width where drifting pebbles sublimate by $16.7$wt\% to $88.3$wt\% from the initial value. The advection-dominated and diffusion-dominated regimes are clearly identified by their different $\Delta x_{\rm subl}$ values in Fig.~\ref{fig_sub_width}.  We explain the details of these different regimes below.

As reference values, $\tau_{\rm s,p} \sim 0.06$, $r_{\rm p} \sim 2$ cm at $r_{\rm snow} \sim 2.4$ au for $F_{\rm p/g}=0.1$ and $\alpha_{\rm acc}=10^{-2}$, while $\tau_{\rm s,p} \sim 0.06$, $r_{\rm p} \sim 20$ cm at $r_{\rm snow} \sim 2.1$ au for $F_{\rm p/g}=0.1$ and $\alpha_{\rm acc}=10^{-3}$. Here, the drag relations correspond to those of the Epstein regime. Below, a notation of ''snow'' indicates values at the snow line.

\subsection{Advection-dominated regime} \label{sec_advection_dominated}

In the advection-dominated regime ($\alpha_{\rm Dr} \ll \alpha_{\rm acc}$), the gas accretion onto the central star dominates the radial transport of vapor. Outside the snow line, the loss of vapor due to the inward advection of the gas balances the supply of vapor from inward drifting icy pebbles. Thus, the icy pebbles locally sublimate to satisfy $P_{\rm sat}(r) \simeq P_{\rm vap}(r)$ outside the snow line ($r > r_{\rm snow}$; see Fig.~\ref{fig_sub_schematic}). Neglecting the effects of the radial diffusion implies that
\begin{align}
	P_{\rm sat}(r) & = P_{\rm vap}(r),
\end{align}
which is rewritten assuming a rapid vertical mixing with the background gas (the molecular weight $\mu_{\rm vap}$) as 
\begin{align}
	P_{\rm 0} \exp \left( - \frac{T_{0}}{T(r)} \right) & = \frac{k_{\rm B} T(r)}{\mu_{\rm vap} m_{\rm proton}} \frac{\Sigma_{\rm vap}(r)}{\sqrt{2\pi} H_{\rm g}(r)} ,
\label{eq_Ps_Pvap}
\end{align}
where $T(r) \propto r^{-\beta}$, $H_{\rm g} \propto r^{(3-\beta)/2}$, and $\nu_{\rm acc}(r) \propto \alpha_{\rm acc} r^{(3-2\beta)/2}$.\\

For $r  \geq r_{\rm snow}$, the surface density of the water vapor that satisfies the local saturation state  is given as
\begin{equation}
\label{eq_vapor}
	\Sigma_{\rm vap}(r) = \frac{\sqrt{2\pi} H_{\rm g}(r) \mu_{\rm vap} m_{\rm proton}}{k_{\rm B} T(r)} P_{\rm 0} \exp \left( -\frac{T_{\rm 0}}{T(r)} \right) .
\end{equation}
Using the surface density of the water vapor and temperature at the snow line (i.e., $\Sigma_{\rm vap,snow} \equiv \Sigma_{\rm vap}(r_{\rm snow})$ and $T_{\rm snow} \equiv T(r_{\rm snow})$, respectively), and Eq.~(\ref{eq_vapor}), $P_{\rm 0}$ is removed and $\Sigma_{\rm vap}(r)$ for $r \geq r_{\rm snow}$ is given as
\begin{align}
	&\Sigma_{\rm vap}(r \geq r_{\rm snow}) \nonumber \\ 
	& = \Sigma_{\rm vap,snow}\frac{T_{\rm snow}}{T(r)} \frac{H_{\rm g}(r)}{H_{\rm g}(r_{\rm snow})} \exp \left( - \frac{T_{\rm 0}}{T(r)} +  \frac{T_{\rm 0}}{T_{\rm snow}} \right) \nonumber \\
	& = \Sigma_{\rm vap,snow} \left( \frac{r}{r_{\rm snow}} \right)^{\beta} \left( \frac{r}{r_{\rm snow}} \right)^{\frac{3-\beta}{2}} \exp \left[ -T_{\rm 0} \left( \frac{1}{T(r)} - \frac{1}{T_{\rm snow}} \right) \right] \nonumber \\
	& = \Sigma_{\rm vap,snow} \left( \frac{r}{r_{\rm snow}} \right)^{\frac{3+\beta}{2}} \exp \left[ \frac{T_{\rm 0}}{T_{\rm snow}} \left( 1 - \left( \frac{r}{r_{\rm snow}} \right)^{\beta} \right) \right]. 
\end{align}
Inside the snow line (i.e., $r \leq r_{\rm snow}$), $\Sigma_{\rm vap}(r)$ is written in terms of the gas mass flux $\dot{M}_{\rm g}$ and the effective viscosity $\nu_{\rm acc}$ as
\begin{equation}
	\Sigma_{\rm vap}(r \leq r_{\rm snow}) = \frac{ \left( 1 - f_{\rm d/p} \right) F_{\rm p/g}\dot{M}_{\rm g}}{3\pi \nu_{\rm acc}(r)} = \frac{\left( 1 - f_{\rm d/p} \right) F_{\rm p/g}\dot{M}_{\rm g}}{3\pi \alpha_{\rm acc}H_{\rm g}^{2}(r)\Omega_{\rm K}(r)}.
\end{equation}
The above equation shows a good match with our 1D simulations for $\alpha_{\rm Dr} \ll \alpha_{\rm acc}$, and thus the advection dominates the radial mass transfer of vapor for such a low-diffusivity case. As discussed above, the local vapor pressure equals the saturation pressure as
\begin{equation}
\label{eq_Pvap_Psat}
	P_{\rm vap}(r) = P_{\rm 0} \exp \left(- \frac{T_{\rm 0}}{T(r)} \right).
\end{equation}
Using the vapor pressure at the snow line $P_{\rm vap,snow} \equiv P(r_{\rm snow})$ and removing $P_{\rm 0}$, the equation is rewritten as
\begin{align}
	\frac{P_{\rm vap}(r)}{P_{\rm vap,snow}} & = \exp \left[ -\frac{T_{\rm 0}}{T(r)} + \frac{T_{\rm 0}}{T_{\rm snow}}  \right] \nonumber \\
		& = \exp \left[ \frac{T_{\rm 0}}{T_{\rm snow}} \left( 1 -  \left( \frac{r}{r_{\rm snow}} \right)^{\beta} \right) \right] .
\end{align}
Using $\Delta x^{*} = (r-r_{\rm snow})/r_{\rm snow} \ll 1$ near the snow line (and $\Delta x^{*} > 0$), the above equation is approximated as follows: 
\begin{align}
	\frac{P_{\rm vap}(r)}{P_{\rm vap,snow}} & =  \exp \left[ \frac{T_{\rm 0}}{T_{\rm snow}} \left( 1 -  \left( \frac{r-r_{\rm snow}}{r_{\rm snow}} + 1 \right)^{\beta} \right) \right] \nonumber \\
	& = \exp \left[ \frac{T_{\rm 0}}{T_{\rm snow}} \left( 1 -  \left( 1+\Delta x^{*}  \right)^{\beta} \right) \right] \\
	& \simeq \exp \left[ \frac{T_{\rm 0}}{T_{\rm snow}} \left( 1 -  \left( 1+ \beta \Delta x^{*}  \right) \right) \right]\\
	& = \exp \left( -\frac{\beta T_{\rm 0}}{T_{\rm snow}} \Delta x^{*} \right) ,
\label{eq_pvap_app}
\end{align}
where $P_{\rm vap}(r)/P_{\rm vap,snow}=0$ is the innermost radius where the sublimation takes place and $P_{\rm vap}(r)/P_{\rm vap,snow}=1$ is the radial location where all ice sublimates (in reality, a very small fraction of pebbles could sublimate inside the snow line). Here, the sublimation width $\Delta x_{\rm subl}$ is equivalent to the radial width of $P_{\rm vap}(r)/P_{\rm vap,snow}=0.167$ to $0.833$, which corresponds to the change in mass of pebbles ($\dot{m}_{\rm p} \propto \dot{\Sigma}_{\rm vap} \propto \dot{P}_{\rm vap}$). Using Eq.~(\ref{eq_pvap_app}), $\Delta x_{\rm subl}^{*} =\Delta x_{\rm subl}/r_{\rm snow}$ is given by 
\begin{align}
	\Delta x_{\rm subl}^{*} & = \Delta x^{*}_{(P_{\rm vap}/P_{\rm vap,snow}=0.167)} - \Delta x^{*}_{(P_{\rm vap}/P_{\rm vap,snow}=0.883)} \nonumber \\
		& = -\frac{T_{\rm snow}}{\beta T_{\rm 0}} \left( \ln{0.167} - \ln{0.883} \right) \\
		& \simeq 0.093 \times \left( \frac{T_{\rm snow}}{170 \, \rm K} \right) \left( \frac{T_{\rm 0}}{6062 \, \rm K} \right)^{-1} \left( \frac{\beta}{0.5} \right)^{-1} ,
\label{eq_x_subl_adv}
\end{align}
where $T_{\rm snow}$ depends on $\alpha_{\rm acc}$ and $F_{\rm p/g}$ (Fig.~\ref{fig_sub_width}). For $r_{\rm snow} \simeq 2.4$ au with $\beta = 0.5$ ($T_{\rm snow} \simeq 170$ K),  $\Delta x_{\rm subl} = \Delta x_{\rm subl}^{*} \times r_{\rm snow} \simeq 0.22$ au ($\simeq 2.2 H_{\rm g}$), which is consistent with the results of 1D simulations (the light-blue line in Fig.~\ref{fig_sub_width} bottom panel). Both 1D simulations and analytical arguments show that the sublimation width in the advection-dominated regime is very weakly dependent on $F_{\rm p/g}$ and $\alpha_{\rm Dr}/\alpha_{\rm acc}$. Equation (\ref{eq_x_subl_adv}) also indicates that the sublimation width in the advection-dominated regime is independent of the Stokes number of pebbles and initial pebble physical size, which are supported by 1D simulations (Fig.~\ref{fig_sub_width}; $r_{\rm p} \sim 2$ cm and $\sim 20$ cm for $\alpha_{\rm acc}=10^{-2}$ and $10^{-3}$, respectively). These are because the sublimation width in the advection-dominated regime is regulated by $P_{\rm sat}$ and $T_{\rm snow}$.

 \subsection{Diffusion-dominated regime}  \label{sec_diffusion_dominated}

In the diffusion-dominated regime ($\alpha_{\rm Dr} \sim \alpha_{\rm acc}$), the local sublimation of drifting pebbles does not take place outside of the snow line ($r > r_{\rm snow}$; see Fig.~\ref{fig_sub_schematic}) because the water vapor produced in the inner region efficiently diffuses outward and the local pressure by the diffused vapor becomes larger than the local saturation vapor pressure, being a super-saturation state (i.e., $P_{\rm vap}(r) > P_{\rm sat}(r)$ for $r > r_{\rm snow}$). Thus, the sublimation only takes place after pebbles pass through the snow line where $P_{\rm vap}(r) < P_{\rm sat}$ (i.e., $r < r_{\rm snow}$).

The solar composition H/He gas sound velocity is given by $c_{\rm s} = \sqrt{k_{\rm B}T/2.34 m_{\rm proton}} \simeq 10^5 (T/280{\, \rm K})^{1/2}$ cm s$^{-1}$. The water vapor sound velocity with the molecular weight $\mu = 18$ is given by $c_{\rm s,v} = \sqrt{k_{\rm B}T/\mu m_{\rm proton}} \simeq 0.36 c_{\rm s} \simeq 0.28 \times 10^4 (T/170{\, \rm K})^{1/2}$ cm s$^{-1}$. Because the saturation pressure and vapor pressure are respectively given by $P_{\rm sat} = c_{\rm s,v}^2 \rho_{\rm sat}$ and $P_{\rm vap} = c_{\rm s,v}^2 \rho_{\rm vap}$ and $m_{\rm p} = (4\pi/3) \rho_{\rm bulk} r_{\rm p}^3$,  where $\rho_{\rm bulk}$ ($\sim 1$ g cm$^{-3}$) is the bulk density of the particle, the rate of change in particle size is given as (see Eqs.~(\ref{eq_subl0}) and (\ref{eq_subl1}))
\begin{equation}
	\frac{dr_{\rm p}}{dt} = - \frac{1}{\sqrt{2\pi} c_{\rm s,v}} 
	\frac{P_{\rm sat}-P_{\rm vap}}{ \rho_{\rm bulk}} .
\label{eq_subl1}
\end{equation}

For the disk density similar to MMSN, the snow line corresponds to $T = T_{\rm snow} \simeq 170$ K. Using Eq.~(\ref{eq_Ps_Pvap}), the vapor pressure with 170 K at $r=r_{\rm snow}$ is $P_{\rm vap,snow} \,[\rm dyn \, cm^{-2}] = 10^{13.06 - 2632/170} \simeq 10^{-2.4} \simeq 3.8 \times 10^{-3}$. We assume that $\Sigma \propto r^{-\beta_\Sigma}$, $T \propto r^{-\beta_T}$, $P_{\rm vap} \propto r^{-\beta_P}$.

We consider the case where the vapor pressure $P_{\rm vap}$ is proportional to disk gas pressure $P_{\rm H/He}$. In that case, $\beta_P = \beta_\Sigma + \beta_T/2 + 3/2$ for ideal gas. We note that $P_{\rm vap} \propto P_{\rm H/He}$ is established in the limit of efficient radial diffusion (i.e., diffusion-dominated regime). For a nominal case, $\beta_\Sigma = 1, \beta_T=1/2$, and $\beta_P = 11/4$.
Inside the snow line ($r < r_{\rm snow}$), we can write
\begin{align}
	& P_{\rm sat}  \simeq 10^{13.06 - \frac{2632\,{\rm K}}{T_{\rm snow}(r/r_{\rm snow})^{-\beta_T}}} = P_{\rm vap,snow} \times 10^{15.5[1-(r/r_{\rm snow})^{\beta_T}]}, 
\label{eq_Psat02} \\
	& P_{\rm vap} \simeq  P_{\rm vap,snow} (r/r_{\rm snow})^{-\beta_P},  
\label{eq_Pv02}
\end{align}
where $T_{\rm snow} \simeq 170$ K was used. At $r \sim r_{\rm snow}$, Eq.~(\ref{eq_subl1}) is reduced to
\begin{align}
	\frac{dr_{\rm p}}{dt} & 
	\simeq \frac{P_{\rm vap}}{\sqrt{2\pi} \, c_{\rm s,v} \, \rho_{\rm bulk}} \left(1- \frac{P_{\rm sat}}{P_{\rm vap}} \right) \nonumber \\
	& \simeq C_P \left(\frac{r}{r_{\rm snow}}\right)^{-\beta_P + \beta_T/2}  \left[ 1 - 10^{15.5(1 - (r/r_{\rm snow})^{\beta_T}]} \left(\frac{r}{r_{\rm snow}}\right)^{\beta_P} \right] c_{\rm s},
\label{eq_subl1B}
\end{align}
where
\begin{align}
	C_P & \equiv \frac{P_{\rm vap,snow}}{\sqrt{2\pi} \, c_{\rm s,v} c_{\rm s}\, \rho_{\rm bulk}} = \frac{P_{\rm vap,snow}}{\sqrt{2\pi} \times 0.36\, c_{\rm s}^2 \, \rho_{\rm bulk}} \nonumber \\
	 & = 0.70 \times 10^{-12} \left(\frac{c_{\rm s,snow}}{0.78 \times 10^5\,\rm cm\,s^{-1}}\right)^{-2} \left(\frac{\rho_{\rm bulk}}{1\,\rm g\,cm^{-3}}\right)^{-1}.
\end{align}
Defining $\Delta \tilde{r} \equiv (r - r_{\rm snow}) / r_{\rm snow}$,
\begin{align}
	\frac{dr_{\rm p}}{dt} & \simeq C_P \left(\frac{r}{r_{\rm snow}}\right)^{-\beta_P + \beta_T/2} \left[ 1 - 10^{-15.5 \beta_T \Delta \tilde{r}} \left(1 + \beta_P \Delta \tilde{r} \right) \right] c_{\rm s} \nonumber \\
	 & \simeq C_P \left( 35.7 \beta_T - \beta_P \right) \Delta \tilde{r} c_{\rm s}.
\label{eq_subl2}
\end{align}
For the nominal parameters, $35.7 \beta_T - \beta_P \simeq 15.1$.
The radial drift of a sublimating pebble is given by
\begin{equation}
	\frac{dr}{dt} = -\Lambda^2 \frac{2 \tau_{\rm s,p}}{1+\Lambda^2 \tau_{\rm s,p}^2}\eta v_{\rm K} + \Lambda \frac{1}{1+\Lambda^2 \tau_{\rm s,p}^2} v_{\rm g,\nu}^{*}.
\label{eq_vr}
\end{equation}
In the case of $\Lambda \simeq 1$ and $\tau_{\rm s,p}^2 \ll 1$,
\begin{equation}
	\frac{d \Delta \tilde{r}}{dt} \simeq - \left( \beta_P \, \tau_{\rm s,p}  + (3/2) \alpha_{\rm acc} \right) \left(\frac{H_{\rm g}}{r}\right) c_{\rm s} \, r_{\rm snow}^{-1}.
\label{eq_vr2}
\end{equation}
From Eqs.~(\ref{eq_subl2}) and (\ref{eq_vr2}) with $\tilde{r}_{\rm p} = r_{\rm p}/r_{\rm p,snow}$,
\begin{align}
	\frac{d \tilde{r}_{\rm p} }{d \Delta \tilde{r}} & \simeq - C_P \frac{ 35.7 \beta_T - \beta_P}{\beta_P \tau_{\rm s,p} + (3/2) \alpha_{\rm acc}} \frac{r_{\rm snow}}{r_{\rm p,snow}}  	\left(\frac{H_{\rm g}}{r}\right)^{-1}_{\rm snow} \Delta \tilde{r} \\	
	 & = -0.98 \times 10^{4} \left(\frac{H_{\rm g}/r}{0.04}\right)^{-1}_{\rm snow} \times \\ 
	 &\left( \frac{\rho_{\rm bulk}}{1\,\rm g\,cm^{-3}} \right)^{-1} \left( \frac{ r_{\rm p,snow}}{1\,\rm cm} \right)^{-1} \left( \frac{ r_{\rm snow}}{2.5\,\rm au} \right) \frac{ 1 }{\beta_P \tau_{\rm s,p} + (3/2) \alpha_{\rm acc}} \Delta \tilde{r}. \nonumber
\label{eq_Rp}
\end{align}
We further set $\tau_{\rm s,p} = \tau_{\rm s,p,snow} \tilde{r}_{\rm p}^\zeta$ (where $\zeta = 2$ for Stokes drag and $\zeta = 1$ for Epstein drag). Then, 
\begin{align}
	\frac{d \tilde{r}_{\rm p} }{d \Delta \tilde{r}} & \simeq \left\{
\begin{array}{ll}
\displaystyle
	-3.6 \times 10^4 C_R \left( \frac{ \tau_{\rm s,p,snow}}{0.1}\right)^{-1} \left( \frac{ r_{\rm p,snow}}{1\,\rm cm} \right)^{-1} \tilde{r}_{\rm p}^{-\zeta} \Delta \tilde{r} \vspace{1em} \\ 
\displaystyle
	  \hspace{7em} \left[ {\rm for \, \,} \tilde{r}_{\rm p} > \tilde{r}_{\rm p,tr} \equiv \left(\frac{\alpha_{\rm acc}}{1.83 \tau_{\rm s,p,snow}}\right)^{1/\zeta} \right] \vspace{1em} \\
\displaystyle
	 -6.6 \times 10^5 C_R \left( \frac{\alpha_{\rm acc}}{10^{-2}} \right)^{-1} \left( \frac{ r_{\rm p,snow}}{1\,\rm cm} \right)^{-1} \Delta \tilde{r} \vspace{1em} \\
\displaystyle
	 \hspace{7em} \left[ {\rm otherwise} \right] 
\end{array}
\right.
\\
C_R & \equiv \left(\frac{H_{\rm g}/r}{0.04}\right)^{-1}_{\rm snow} \left( \frac{\rho_{\rm bulk}}{1 {\,\rm g \,cm^{-3}}} \right)^{-1} \left( \frac{ r_{\rm snow}}{2.5 {\,\rm au}} \right).
\end{align}
We integrate the upper equation from $\tilde{r}_{\rm p}=1$ as
\begin{align}
	\tilde{r}_{\rm p} & (\Delta \tilde{r}) \simeq \nonumber \\
	& \left[1 - 3.6 \times 10^4 C_R \left(\frac{\zeta+1}{2}\right) \left( \frac{ \tau_{\rm s,p,snow}}{0.1}\right)^{-1} \left( \frac{ r_{\rm p,snow}}{1\,\rm cm} \right)^{-1} \Delta \tilde{r}^2 \right]^{1/(\zeta+1)}.
\end{align}
If we define the sublimation width $\Delta x_{\rm subl}$ by the radial separation between $\tilde{r}^3_{\rm p}=\xi_1$ and $\tilde{r}^3_{\rm p}=\xi_2$ (as $m_{\rm p} \propto \tilde{r}^3_{\rm p}$),
\begin{align}
  \Delta & x_{\rm subl}  \simeq  \left[ \left( 1 - \xi_1^{(\zeta+1)/3} \right)^{1/2} - \left( 1-\xi_2^{(\zeta+1)/3} \right)^{1/2} \right] \times \nonumber \\
  & 0.13 C_R^{-1/2} \left(\frac{\zeta+1}{2} \right)^{-1/2} \left(\frac{H_{\rm g}/r}{0.04}\right)^{-1} \left(\frac{\tau_{\rm s,p,snow}}{0.1}\right)^{1/2} \left( \frac{ r_{\rm p,snow}}{1\,\rm cm} \right)^{1/2} H_{\rm g}.
\label{eq_Dr3}
\end{align}

Equation (\ref{eq_Dr3}) (with $\zeta=1$, $\tau_{\rm s,p}=0.06$, $r_{\rm p}=2.3$ cm, $r_{\rm snow}=2.4$ au, $\xi_1=0.167$, and $\xi_2=0.883$) is shown by the light-green line in the bottom panel of Fig.~\ref{fig_sub_width}. The analytical arguments above are generally in accordance with the 1D numerical results. The sublimation width in the diffusion-dominated regime depends on $F_{\rm p/g}$, in that, a larger $F_{\rm p/g}$ leads to a smaller $\Delta x_{\rm subl}$. This is because the location of the snow line becomes closer to the star for a larger $F_{\rm p/g}$, which leads to a larger $P_{\rm sat}-P_{\rm vap}$ (see Eq.~(\ref{eq_subl1})). In the case of $\alpha_{\rm acc}=10^{-3}$ (squares in Fig.~\ref{fig_sub_width}), the snow line is located closer to the star (a larger $P_{\rm sat}-P_{\rm vap}$) than that of $\alpha_{\rm acc}=10^{-2}$, which makes $\Delta x_{\rm subl}$ smaller, whereas the size of pebbles is 10 times larger than that of $\alpha_{\rm acc}=10^{-2}$, which makes $\Delta x_{\rm subl}$ larger.

\subsection{Boundary between advection/diffusion-dominated regimes} \label{sec_boundary}

The boundary between two regimes $-$ the advection-dominated regime and the diffusion-dominated regime $-$ can be evaluated by considering the mass fluxes of advection and diffusion. In the steady-state for an unperturbed case (no back-reaction), the rate of change in the surface density of the vapor (Eq.~(\ref{eq_sigma_vap})) is described as: 
\begin{align}
	& \frac{1}{r}\frac{\partial}{\partial r}  \left( r \Sigma_{\rm vap} v_{\rm g,\nu} - r D_{\rm Dr,g}\Sigma_{\rm g}  \frac{\partial}{\partial r}  \left( \frac{\Sigma_{\rm vap}}{\Sigma_{\rm g} } \right) \right) \nonumber  \\ 
	& = \frac{1}{r}\frac{\partial}{\partial r}  \left( r \Sigma_{\rm vap} v_{\rm g,\nu} -  D_{\rm Dr,g}  \Sigma_{\rm vap} \frac{\partial \ln \Sigma_{\rm vap}}{\partial \ln r} \right) \nonumber \\
	& = \frac{1}{r}\frac{\partial}{\partial r}  \left( r \Sigma_{\rm vap} v_{\rm g,\nu} -  D_{\rm Dr,g}  \Sigma_{\rm vap} \frac{\partial \ln P_{\rm vap}}{\partial \ln r} \right) ,
	\label{eq_vap_flux}
\end{align}
where we assume that the change in the water vapor is more rapid than the other quantities (i.e., we only consider the local change in the water vapor and $\partial P_{\rm vap}/\partial r \simeq \partial \Sigma_{\rm vap}/\partial r$). As discussed in the previous subsection (Section \ref{sec_advection_dominated}), $P_{\rm vap}$ satisfies the local saturation vapor (Eq.~(\ref{eq_Pvap_Psat})) and 
\begin{equation}
	\frac{\partial \ln P_{\rm vap}}{\partial \ln r} = - \beta \frac{T_{\rm 0}}{T(r)}.
\end{equation}
From Eq.~(\ref{eq_vap_flux}), the diffusion dominates over the advection when 
\begin{align}
	 & r \Sigma_{\rm vap} v_{\rm g,\nu} -  D_{\rm Dr,g}  \Sigma_{\rm vap} \frac{\partial \ln P_{\rm vap}}{\partial \ln r} > 0,
\end{align}
which is rewritten as
\begin{align}
	 & c_{\rm s}^2\Omega_{\rm K}^{-1} \Sigma_{\rm vap} \left( -\frac{3}{2} \alpha_{\rm acc} + \beta \frac{T_{\rm 0}}{T(r)} \alpha_{\rm Dr} \right) > 0 .
\end{align}
Therefore, the diffusion dominates over the advection when 
\begin{equation}
	\frac{\alpha_{\rm Dr}}{\alpha_{\rm acc}} > \frac{3}{2\beta} \frac{T(r)}{T_{\rm 0}} \simeq 0.08 \times \left( \frac{\beta}{0.5} \right)^{-1} \left( \frac{T_{\rm 0}}{6062 \,\rm K} \right)^{-1} \left( \frac{T(r)}{170 \,\rm K} \right).
\label{eq_boundary}
\end{equation}
The analytical prediction of Eq.~(\ref{eq_boundary}) (with $r=2.4$ au for $\beta=0.5$) is shown by a gray dashed line in Fig.~\ref{fig_sub_width} and it shows a good consistency with 1D numerical simulations.

\subsection{A quick summary of sublimation width of drifting pebbles} \label{sec_fitting}
As discussed above, the sublimation width of icy drifting pebbles is categorized by two regimes $-$ the advection-dominated regime (Section \ref{sec_advection_dominated}) and the diffusion-dominated regime (Section \ref{sec_diffusion_dominated}) $-$ which is a function of $\alpha_{\rm Dr}/\alpha_{\rm acc}$. In the lower and higher $\alpha_{\rm Dr}/\alpha_{\rm acc}$ regions, advection and radial diffusion are dominated, respectively. Here, we provide a fitting function that smoothly connects the two regimes as 
\begin{align}
\label{eq_fitting}
\log_{10} \left(\frac{\Delta x_{\rm subl}}{H_{\rm g}} \right) 
& \simeq \frac{X_+ - X_-}{2} 
{\rm erf} 
\left[ 3 \log_{10} \left( \frac{ \alpha_{\rm Dr}/\alpha_{\rm acc}}{0.08} \right) \right] + \frac{X_+ + X_-}{2}, \\
 & {\rm where \,\,} X_- = \log_{10}( 2 )  {\rm \,\, and \,\,}  X_+ = \log_{10}(0.1). \nonumber
\end{align} 
Equation (\ref{eq_fitting}) is shown by a black line in Fig.~\ref{fig_sub_width} and is used to characterize $H_{\rm d}(\Delta x_{\rm subl})$ in Section \ref{sec_results}.

\subsection{Effects of sintering on the sublimation of pebbles}
When sintering effects are considered \citep[e.g.,][]{Sai11,Oku16}, a sudden destruction of pebbles may occur during the sublimation and the resultant sublimation width may become smaller than those considered above (Sections \ref{sec_advection_dominated} and \ref{sec_diffusion_dominated}), where a gradual decrease in size is expected. Pebbles would sublimate by a certain fraction to produce enough vapor that triggers a sudden destruction of pebbles due to sintering effects. Thus, the sublimation width may be reduced by some fraction from the above arguments when sintering is taken into account.
This 
would lead to a higher midplane concentration of silicate dust (smaller $\Delta x_{\rm subl}$ leads to a smaller $H_{\rm d}$). Details of the effects of sintering on the sublimation width of drifting pebbles are beyond the scope of this paper. We leave this matter for future investigations.

\section{Scale height and maximum flux of pebbles}
\label{sec_KH_ND}

\subsection{Pebble scale height regulated by KH instability} \label{sec_KH}

When the midplane concentration of pebbles/dust increases, a vertical shear in the gas velocity between the pebble-concentrated layer and the upper layer induces vorticity at the interface between the two fluids, i.e., Kelvin-Helmholtz (KH) instability \citep[e.g.,][]{Sek98,You02,Chi08}.

The azimuthal component of gas velocity \citep{Sch17} is  
\begin{align}
	 v_{\rm g,\phi} & = -\Lambda \frac{1+\Lambda\tau_{\rm s}^2}{1+\Lambda^2 \tau_{\rm s}^2} \eta v_{\rm K} - Z \frac{\frac{1}{2}\Lambda^2\tau_{\rm s}}{1+\Lambda^2 \tau_{\rm s}^2} v_{\rm g,\nu}^{*}\\
	& \simeq - \frac{\Lambda}{1+\Lambda^2 \tau_{\rm s}^2} \left( \left(1+\Lambda \tau_{\rm s}^2 \right) - \frac{3}{4} \frac{Z\tau_{\rm s}}{1+Z} \alpha_{\rm acc} C_{\rm \eta}^{-1} \right) \eta v_{\rm K} .
\end{align}
Because $\tau_{\rm s} \ll 1$, $\alpha_{\rm acc} \ll 1$, and $|v_{\rm g,\phi}| \gg |v_{\rm g,r}|$, $v_{\rm g,\phi}$ is approximated as
\begin{equation}
	v_{\rm g,\phi} \simeq -\Lambda \eta v_{\rm K} = \frac{1}{1+Z} \eta v_{\rm K}.
\end{equation}

The Richardson number for pebbles at vertical height $z$ is 
\begin{align}
	 Ri & = - \frac{g \left( d\left( \rho_{\rm g} + \rho_{\rm p} \right)/dz \right)}{ \left( \rho_{\rm g} + \rho_{\rm p} \right) \left( dv_{\rm g,\phi}/dz \right)^2} \nonumber \\
	& \simeq \frac{\Omega_{\rm K}^2 z \times \left( \rho_{\rm g} H_{\rm g} ^{-2} + \rho_{\rm p} H_{\rm p} ^{-2} \right)z }{ \left( \rho_{\rm g} + \rho_{\rm p} \right) \left( \frac{Z}{(1+Z)^2} \eta v_{\rm K} z H_{\rm p}^{-2} \left( 1- h_{\rm p/g}^2 \right)  \right)^2 }\\
	& \simeq \frac{ \left( 1+Zh_{\rm p/g}^{-2} \right) \Omega_{\rm K}^2 \left( z/H_{\rm g} \right)^2 }{ \frac{Z^2}{\left( 1+Z \right)^3} \eta^2 \Omega_{\rm K}^2 \left( rz/H_{\rm p}^2 \right)^2 \left( 1- h_{\rm p/g}^2 \right)^2 }\\
	& \simeq \frac{ \left( 1+Z \right)^3}{Z^2} C_{\rm \eta}^{-2} \left( \frac{H_{\rm g}}{r} \right)^{-2} \left( 1 + Zh_{\rm p/g}^{-2} \right) \left( \frac{h_{\rm p/g}^2}{1-h_{\rm p/g}^2} \right)^2 ,
\label{eq_Ri}
\end{align}
where assumptions are made for $\rho_{\rm g}(z) \propto \exp \left( -z^2/2H_{\rm g}^2 \right)$, $\rho_{\rm p}(z) \propto \exp \left( -z^2/2H_{\rm p}^2 \right)$, and $g \simeq \Omega_{\rm K}^2z$. We define $C$ as 
\begin{equation}
	C \equiv Ri \frac{Z^2}{ \left(1+Z \right)^3 } C_{\rm \eta}^2 \left( \frac{H_{\rm g}}{r} \right)^2 
\end{equation}
Defining $X \equiv h_{\rm p/g}^2$ and using $C$ and $X$, we can rewrite Eq.~(\ref{eq_Ri}) as
\begin{equation}
	 C = \frac{ \left(X+Z\right)X}{\left( 1-X \right)^2},
\end{equation}
which is equivalently written as
\begin{equation}
	 \left( 1-C \right)X^2 + \left( Z+2C \right)X - C = 0.
\end{equation}
As $Z \gg C$ by definition and $X \ll 1$, the solution is $X \simeq C/Z$, that is, the scale height of pebbles regulated by a KH instability is written as 
\begin{align}
	 h_{\rm p/g,KH} & \simeq Ri^{1/2} \frac{Z^{1/2}}{ \left( 1+Z \right)^{3/2} } C_{\rm \eta} \left( \frac{H_{\rm g}}{r} \right)\\
	 & \simeq  0.039 \left(\frac{Ri}{0.5}\right)^{1/2} \, \frac{Z^{1/2}}{(1+Z)^{3/2}}  \left( \frac{C_{\rm \eta}}{11/8} \right) \left(\frac{H_{\rm g}/r}{0.04}\right) , 
\label{eq_hpg_KH}
\end{align}
where Eq.~(\ref{eq_hpg_KH}) has a peak value at $Z={1/2}$. We note that Eq.~(\ref{eq_hpg_KH}) with $Ri=0.5$ and $Z=0.5$ ($h_{\rm p/g} \equiv H_{\rm p}/H_{\rm g}$)  is equivalent, within a factor order unity, to a classical description of $H_{\rm p} \propto \eta r$ \citep[e.g.,][]{Chi10}. Although $Ri$ has a dependence on $Z$ and $Z>1$ could lead to SI and/or cusp \citep{Sek98,You02}, the average $Ri \sim 0.75$ is reported for $Z<1$ (or $Z_{\rm \Sigma} < 0.01$) \citep{Ger20}. Thus, our simple prescription is valid for $Z<0$ (this is where we use the above prescription to derive a critical $H_{\rm p}$). Further detailed discussion with numerical simulations is beyond the scope of this paper.

We note also that we estimated $h_{\rm p/g}$, assuming a Gaussian distribution for heights of the particles,
while the precise height distribution function has a cusp for $Z \ga 1$ and the cusp becomes sharper as $Z$ increases \citep{Sek98,Chi08}. The height of the cusp ($z_{\rm max}$) at which the particle distribution is truncated becomes constant in the limit of $Z \gg 1$ \citep{Chi08}, while Eq.~(\ref{eq_hpg_KH}) shows $h_{\rm p/g,KH} \propto Z^{-1}$ for $Z \gg 1$. On the other hand, the precise distribution is more similar to the Gaussian distribution for $Z \ll 1$, Actually, in this case, $z_{\rm max}$ and $H_{\rm p}$ given by Eq.~(\ref{eq_hpg_KH}) are similar (only differ by $\sqrt{2}$). Because we define $H_{\rm p}$ as a root mean square of heights of the particle distribution, the derivation assuming a Gaussian distribution is more appropriate than the estimation of $H_{\rm p}=z_{\rm max}$.

\begin{figure*}[h]
	\centering
	\resizebox{\hsize}{!}{ \includegraphics{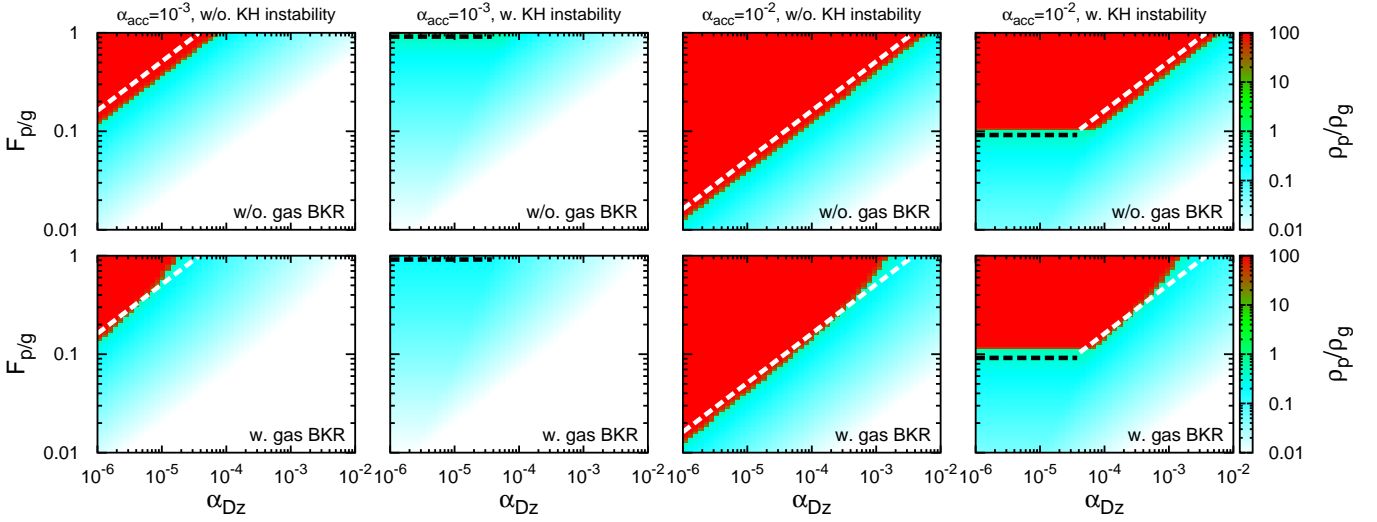} }
	\caption{Midplane pebble-to-gas density ratio ($Z=\rho_{\rm p}/\rho_{\rm g}$) as a function of $\alpha_{\rm Dz}$ and $F_{\rm p/g}$ (at $r=5$ au and $\tau_{\rm s,p}=0.125$). The top panels correspond to the cases where the back-reaction onto the gas motion is neglected (i.e., $v_{\rm g}=v_{\rm g,\nu}$). The bottom panels correspond to the cases where the back-reaction to the gas motion is included, following Eq.~(\ref{eq_vg}). The color contours are obtained by directly solving Eq.~(\ref{eq_peb_pileup}). The left and right two panels are the cases where $\alpha_{\rm acc}=10^{-3}$ and $\alpha_{\rm acc}=10^{-2}$ with/without a KH instability ($Ri=0.5$), respectively. The red-colored regions indicate where a steady-state solution is not found (i.e., $\rho_{\rm p}/\rho_{\rm g}>100$ and it keeps increasing; the "no-drift (ND)" region). The white (Eq.~(\ref{eq_Fpg_cri_regime1})) and black (Eq.~(\ref{eq_Fpg_cri_regime2}) with $Ri=0.5$) dashed lines are analytically derived critical $F_{\rm p/g,crit}$ above which no steady-state are found (the "no-drift (ND)" region). Analytical predictions (dashed lines) well predict critical boundaries of the ND region. Including a KH instability reduces the ND region as the minimum scale height of pebbles is regulated by a KH instability. Due to the back-reaction to the gas motion that slows down the radial velocity of the gas, the midplane concentration of pebbles as well as the ND region (red-colored region) are shifted toward slightly larger $F_{\rm p/g}$ values.}
\label{fig_Fpg_cri}
\end{figure*}

\subsection{"No-drift (ND)" region} \label{sec_ND}

Drift-BKR reduces the radial drift velocity of pebbles as a local concentration of pebbles is elevated \citep[see also][]{Gon17}. The midplane concentration of pebbles increases with increasing $F_{\rm p/g}$ and decreasing $H_{\rm p}$. 

For sufficiently large $F_{\rm p/g}$ and small $H_{\rm p}$ (i.e., small $\alpha_{\rm Dz}$) values, the drift velocity of pebbles can decrease in a self-induced manner due to Drift-BKR, eventually leading to a complete stop of the radial motion of pebbles. In such a case, pebbles no longer reach the snow line from the outer region of the disk. We call this parameter regime as the "no-drift (ND)" region. As shown below, we explain that the ND region appears at an arbitrary choice of the radial distance, i.e., irrespective of the vicinity of the snow line.

Below, we discuss the concentration of icy pebbles in the disk midplane considering Drift-BKR onto pebbles and gas. We numerically and analytically derive the degree of the midplane concentration of pebbles. The results are also used for our initial conditions at the outer boundary of 1D simulations. We identify the ND region (within the $F_{\rm p/g} - \alpha_{\rm acc} - \alpha_{\rm Dz}$ space).

The concentration of pebbles at the midplane is written as
\begin{equation}
	Z \equiv \frac{\rho_{\rm p}}{\rho_{\rm g}} = \frac{\Sigma_{\rm p}}{\Sigma_{\rm g}} h_{\rm p/g}^{-1} = \frac{v_{\rm g}}{v_{\rm p}} h_{\rm p/g}^{-1} F_{\rm p/g} ,
\label{eq_peb_pileup}
\end{equation}
where $v_{\rm g}$, $v_{\rm p}$, and $h_{\rm p/g}$ are functions of $\Lambda(Z)$ and $Z=\rho_{\rm p}/\rho_{\rm g}$ (Eqs.~(\ref{eq_vg}), (\ref{eq_vp}), (\ref{eq_Hp}), and (\ref{eq_lambda})). We numerically find a solution (i.e., solve for $Z$ with $\tau_{\rm s,p}=0.1$) for Eq.~(\ref{eq_peb_pileup}) for a given $F_{\rm p/g}$ and $\alpha_{\rm Dz}$. The color contours in Fig.~\ref{fig_Fpg_cri} show the numerically obtained $\rho_{\rm p}/\rho_{\rm g}$ values. The red-colored regions correspond to the parameters for which no steady-state solution is found (numerically, this corresponds to $\rho_{\rm p}/\rho_{\rm g} \gg 100$), defined as the ND region for which pebbles no longer reach the snow line. We found that the boundaries of the "no-drift" regimes are characterized by a critical value of $\rho_{\rm p}/\rho_{\rm g}=1$ (i.e., $Z_{\rm cri}=1$; see the color contours in Fig.~\ref{fig_Fpg_cri}). The dashed lines in Fig.~\ref{fig_Fpg_cri} are analytically derived critical $F_{\rm p/g,crit}$ above which no steady-state solution is found for $\rho_{\rm p}/\rho_{\rm g}$ (i.e., no solution for $Z$). The analytical estimations well predict the direct solutions (the color contours). Below, we discuss that $F_{\rm p/g,crit}$ can be divided into two different regimes.

The first regime is when $H_{\rm p,tur} > H_{\rm p,KH,max}$ where $H_{\rm p,KH,max}$ is the maximum scale height of pebbles regulated by a KH-instability (Section \ref{sec_KH}). In this case, using Eqs.~(\ref{eq_vg}) and (\ref{eq_vp}), the vertically averaged metallicity of pebbles is given as
\begin{align}
\label{eq_Z_peb}
	Z_{\rm \Sigma}  & \equiv  \frac{\Sigma_{\rm p}}{\Sigma_{\rm g}} = F_{\rm p/g} \times \frac{v_{\rm g}}{v_{\rm p}} \\
	  & \simeq  F_{\rm p/g} \times \nonumber \\
\label{eq_Z_peb_app}
	  & \frac{ \left(\frac{\Lambda}{1+\Lambda^{2}\tau_{\rm s,p}^{2}} \right) \left[ \left( \frac{2\tau_{\rm s,p}}{1+Z} + \frac{3\alpha_{\rm acc}}{2C_{\rm \eta}} \right) Zh_{\rm p/g} - \frac{3\alpha_{\rm acc}}{2C_{\rm \eta}} \left( \frac{1+\Lambda^2 \tau_{\rm s,p}^2}{\Lambda}\right) \right] \eta v_{\rm K}}    { -  \left(\frac{\Lambda}{1+\Lambda^{2}\tau_{\rm s,p}^{2}} \right) \left( 2\Lambda \tau_{\rm s,p} + \frac{3\alpha_{\rm acc}}{2C_{\rm \eta}}  \right) \eta v_{\rm K}}\\
	& \simeq \frac{3 \alpha_{\rm acc} F_{\rm p/g}}{4 \tau_{\rm s,p} C_{\rm \eta} \Lambda^{2}} = \frac{3 \alpha_{\rm acc} F_{\rm p/g}}{4 \tau_{\rm s,p} C_{\rm \eta}} \left( 1 + h_{\rm p/g}^{-1} Z_{\rm \Sigma} \right)^{2} ,
\end{align}
where approximation is made for $\alpha_{\rm acc} \ll \tau_{\rm s,p} \ll 1$, $Z \ll 1$, $\Lambda \simeq 1$ (because $\Lambda= 1 \rightarrow 0$ as pile-up proceeds), and we use Eq.~(\ref{eq_lambda_inv}). Solving this equation gives
\begin{equation}
	Z_{\rm \Sigma} = \frac{\left(1 - 2ab \right) \pm \sqrt{1-4ab}}{2ab^{2}} ,
\end{equation}
where $a=\frac{3\alpha_{\rm acc}F_{\rm p/g} }{4\tau_{\rm s,p} C_{\rm \eta}}$ and $b=h_{\rm p/g}^{-1}$, respectively. $Z_{\rm \Sigma}$ has a real solution when $1-4ab > 0$ and a critical $F_{\rm p/g,crit1}$ for the first regime is given as
\begin{align}
\label{eq_Fpg_cri_regime1}
	 F_{\rm p/g, crit1}  & = \frac{C_{\rm \eta}}{3} \frac{\tau_{\rm s,p}}{\alpha_{\rm acc}} h_{\rm p/g}
	  \simeq \frac{ \left( \alpha_{\rm Dz}\tau_{\rm s,p} \right)^{1/2} C_{\rm \eta}}{3\alpha_{\rm acc}} \\
	 & \simeq 0.15 \times \left( \frac{\alpha_{\rm acc}}{10^{-2}}\right)^{-1} \left( \frac{\alpha_{\rm Dz}}{10^{-4}} \right)^{1/2} \left( \frac{\tau_{\rm s,p}}{0.1} \right)^{1/2}  \left( \frac{C_{\rm \eta}}{11/8} \right), \\
	 & \hspace{11em}  \left[  {\rm  for \,} H_{\rm p,tur} > H_{\rm p,KH,max} \right]  \nonumber
\end{align}
where $h_{\rm p/g} \simeq \left( \tau_{\rm s,p}/\left( \alpha_{\rm Dz} (1+Z)^{-K} \right) \right)^{-1/2} \simeq \left( \tau_{\rm s,p}/\alpha_{\rm Dz} \right)^{-1/2}$ for $Z \ll 1$ and $\alpha_{\rm Dz} \ll \tau_{\rm s,p}$ (Eq.~(\ref{eq_Hp})). The white dashed line in Fig.~\ref{fig_Fpg_cri} shows Eq.~(\ref{eq_Fpg_cri_regime1}) and it generally shows a good consistency with the direct solutions of Eq.~(\ref{eq_peb_pileup}) (i.e., color contours in Fig.~\ref{fig_Fpg_cri})\footnote{Even when the back-reaction to the gas motion is neglected \citep[i.e., $v_{\rm g}=v_{\rm g,\nu}$ as][]{Ida16,Sch17,Dra17,Hyo19}, $F_{\rm p/g, crit1}$ is approximated to $\sim \frac{C_{\rm \eta}}{3} \frac{\tau_{\rm s,p}}{\alpha_{\rm acc}} h_{\rm p/g}$ (i.e., replace the denominator of Eq.~(\ref{eq_Z_peb_app}) with $v_{\rm g,\nu}$ and approximation is made for $\alpha_{\rm acc} \ll \tau_{\rm s,p} \ll 1$, $Z \ll 1$, $\Lambda \simeq 1$). }.

The second regime appears when a KH instability plays a role, which corresponds to $H_{\rm p,tur} < H_{\rm p,KH,max}$, where $H_{\rm p,KH,max}$ is the maximum scale height of pebbles regulated by a KH instability (see Section \ref{sec_KH}). Thus, in this case, the scale height of pebbles is described by $H_{\rm p,KH}$, which is independent of $\alpha_{\rm Dz}$. As $Z=\rho_{\rm p}/\rho_{\rm g}$ increases from zero value, $H_{\rm p,KH}$ initially increases and reaches its maximum $H_{\rm p,KH,max}$ at $Z = 1/2$ (see Section \ref{sec_KH}). As $Z$ further increases, $H_{\rm p,KH}$ decreases. Using $Z_{\rm cri}=1$ as a critical value, the critical $F_{\rm p/g}$ in the second regime $F_{\rm p/g,crit2}$ adopts $H^{Z=1}_{\rm p,KH}$ ($H_{\rm p,KH}$ with $Z_{\rm cri} = 1$). $Z_{\rm cri}$ is given as
\begin{align}
	 Z_{\rm cri} & = \frac{\rho_{\rm p}}{\rho_{\rm g}} \mid_{H_{\rm p,KH}^{Z=1}}
	   = \frac{\Sigma_{\rm p}}{\Sigma_{\rm g}} \left( h_{\rm p/g,KH}^{Z=1} \right)^{-1} \\
	  & \simeq  \frac{3 \alpha_{\rm acc} F_{\rm p/g,crit2}}{4 \tau_{\rm s,p} C_{\rm \eta} (\Lambda_{\rm KH}^{Z=1})^2} \left( h_{\rm p/g,KH}^{Z=1} \right)^{-1} ,
\end{align}
where $h_{\rm p/g,KH}^{Z=1} \equiv H_{\rm p,KH}^{Z=1}/H_{\rm g}$ and approximation is made for $\alpha_{\rm acc} \ll \tau_{\rm s,p} \ll 1$ and $\Lambda \simeq 1$. $\Lambda_{\rm KH}^{Z=1}$ is $\Lambda$ using $H_{\rm p,KH}^{Z=1}$. Here, $Z = 1$ and $\Lambda_{\rm KH}^{Z=1} = 1/2$. Thus, the critical $F_{\rm p/g}$ for the second regime $F_{\rm p/g,crit2}$ with $C_{\rm \eta}=11/8$ for $T \propto r^{-1/2}$ is given as 
\begin{align}
\label{eq_Fpg_cri_regime2}
	 F_{\rm p/g,crit2} & = \frac{C_{\rm \eta}}{3} \left( \frac{\tau_{\rm s,p}}{\alpha_{\rm acc}} \right) h_{\rm p/g,KH}^{Z=1} = \frac{11}{24} \left( \frac{\tau_{\rm s,p}}{\alpha_{\rm acc}} \right) h_{\rm p/g,KH}^{Z=1} \\
	&\simeq 0.06 \times \left( \frac{\alpha_{\rm acc}}{10^{-2}}\right)^{-1} \left( \frac{\tau_{\rm s,p}}{0.1}\right) \left( \frac{Ri}{0.5}\right)^{1/2} \left( \frac{H_{\rm g}/r}{0.04} \right) , \\
	  & \hspace{11em}  \left[  {\rm  for \,} H_{\rm p,tur} < H_{\rm p,KH,max} \right]  \nonumber
\end{align}
which is independent on $\alpha_{\rm Dz}$. Equation (\ref{eq_Fpg_cri_regime2}) is plotted by the black dashed line in Fig.~\ref{fig_Fpg_cri} and it shows a good consistency with the direct solution of Eq.~(\ref{eq_peb_pileup}). Including a KH instability (i.e., the second regime and Eq.~(\ref{eq_Fpg_cri_regime2})) prevents the scale height of pebbles becomes further smaller as $\alpha_{\rm Dz}$ becomes small, reducing the parameter range of the "no-drift" region (see Fig.~\ref{fig_Fpg_cri} for the cases with/without a KH instability).

The above arguments are irrespective of the vicinity of the snow line and the ND mode could occur at an arbitrary radial location to the central star. The application of the "no-drift" mechanism in protoplanetary disks is discussed in \cite{Hyo21}.

\begin{figure*}[t]
	\centering
	\resizebox{\hsize}{!}{ \includegraphics{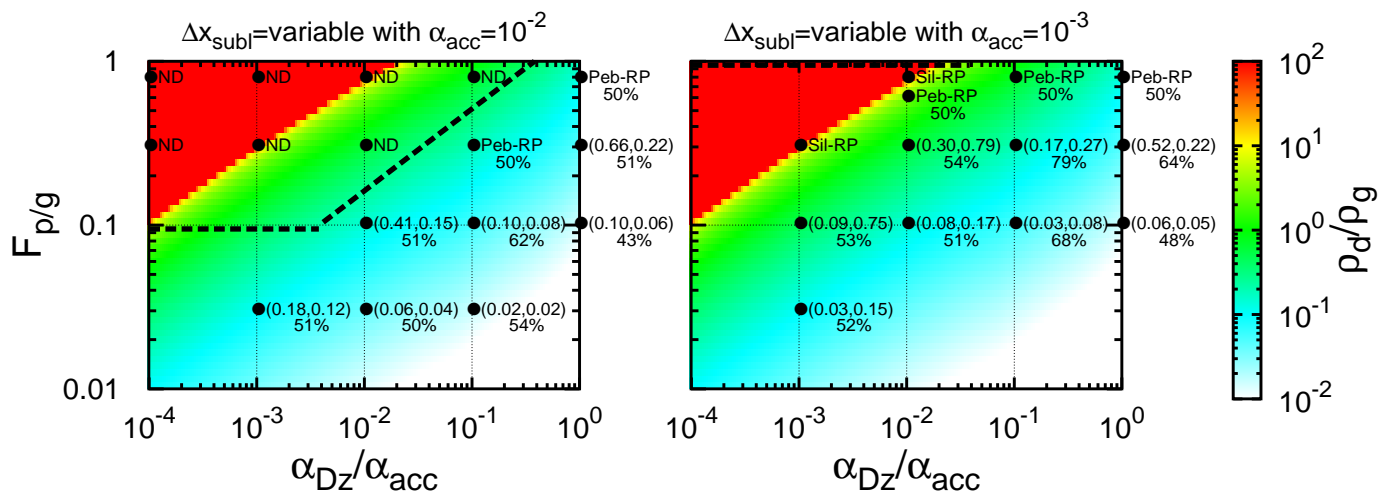} }
	\caption{Zones of pile-up of pebbles and/or silicate dust around the snow line as a function of $F_{\rm p/g}$ and $\alpha_{\rm Dz}(=\alpha_{\rm Dr})/\alpha_{\rm acc}$ (case of a variable $\Delta x_{\rm subl}$; Eq.~(\ref{eq_fitting}) in Section \ref{sec_fitting}). The numbers in the parenthesis represent the maximum $\rho_{\rm p}/\rho_{\rm g}$ (left) and $\rho_{\rm d}/\rho_{\rm g}$ (right) obtained from our full 1D simulations, respectively. The numbers below the parenthesis represent the rock fraction in the surface density of pebbles at the maximum $\rho_{\rm p}/\rho_{\rm g}$ (when a runaway pile-up of silicate dust occurs, the fraction is not shown because the system would gravitationally collapse to form 100\% rocky planetesimals). The runaway pile-up of pebbles (labeled as "Peb-RP") or that of silicate dust (labeled as "Sil-RP") occurs around the snow line, depending on the combinations of $F_{\rm p/g}$ and $\alpha_{\rm Dz}(=\alpha_{\rm Dr}$). The color contour shows an analytical prediction of the maximum $\rho_{\rm d}/\rho_{\rm g}$ just inside the snow line obtained from Paper I. The black dashed line indicates analytically derived critical $F_{\rm p/g}$ above which the "no-drift (ND)" takes place (labeled by "ND" for the results of 1D simulations in the panels) at the outer boundary ($r=5$ au and $\tau_{\rm s,p}=0.125$; Section \ref{sec_ND}). Drift-BKR and Diff-BKR onto the motions of pebbles and silicate dust are included with $K=1$. Left and right panels show the cases of $\alpha_{\rm acc}=10^{-2}$ and $\alpha_{\rm acc}=10^{-3}$, respectively. In the case of $\alpha_{\rm acc}=10^{-2}$, the runaway pile-up of silicate dust is inhibited because the required parameter regime is overlapped by the "no-drift" region (i.e., pebbles do not reach the snow line). In the case of $\alpha_{\rm acc}=10^{-3}$, the runaway pile-ups of both pebbles and silicate dust occur.}
\label{fig_map_variable}
\end{figure*}

\section{Pile-ups around the snow line: Dust or pebbles?} \label{sec_results}

In this section, we finally show the overall results of our full 1D simulations that include sublimation of ice, the release of silicate dust, and the recycling processes (water vapor as well as silicate dust via the recondensation and sticking onto pebbles outside the snow line -- see Fig.~\ref{fig_summary}). We adopt a realistic scale height of silicate dust obtained from Paper I. We also consider the effects of a KH instability for the scale height of pebbles (Section \ref{sec_KH}), which was not included in the previous works. We  include Drift-BKR and Diff-BKR onto the motions of pebbles and silicate dust. We choose $K=1$ for Diff-BKR. We change $F_{\rm p/g}$, $\alpha_{\rm acc}$, and $\alpha_{\rm Dr}=\alpha_{\rm Dz}$ as parameters. For simplicity and to have the same setting as Paper I, we fix the gas structure as an unperturbed background disk (i.e., $v_{\rm g} = v_{\rm g,\nu}^{*}$). 

Figure \ref{fig_map_variable} shows the maximum midplane solid-to-gas ratio around the water snow line in the $F_{\rm p/g}$--$\alpha_{\rm acc}$--$\alpha_{\rm Dz}$(=$\alpha_{\rm Dr}$) space. The background color contours correspond to the results obtained for silicate dust inside the snow line in Paper I. The red-colored area indicates runaway pile-up of silicate dust (i.e., no steady-state and an indication of planetesimal formation via direct gravitational collapse). The dependence on $\Delta x_{\rm sub}$ is discussed in Appendix \ref{sec_dep_subl}.

Points in Fig.~\ref{fig_map_variable} represent the results of our 1D simulations. The left and right numbers in the parenthesis represent the maximum $\rho_{\rm p}/\rho_{\rm g}$ (pebbles) and $\rho_{\rm d}/\rho_{\rm g}$ (silicate dust), respectively. When a runaway pile-up occurs in our 1D simulations, it is labeled either by "Peb-RP" for pebbles or by "Sil-RP" for silicate dust. The numbers below the parenthesis in Fig.~\ref{fig_map_variable} show the rock fraction in pebbles. For a combination of a small $\alpha_{\rm Dz}$(=$\alpha_{\rm Dr}$)/$\alpha_{\rm acc}$ and a large $F_{\rm p/g}$ as an initial setting (as we set pebbles initially exist only at the outer boundary), pebbles do not drift from the outer boundary as a consequence of continuous slowing down of the radial drift velocity of pebbles due to Drift-BKR (we label these parameters by "ND" as the "no-drift" region in Fig.~\ref{fig_map_variable} -- see also Section~\ref{sec_ND}).

In the following subsections, we discuss the comparison to previous works (Section \ref{sec_comparison}), the detailed processes of pile-up of silicate dust (Section \ref{sec_pile-up_dust}), that of pebbles (Section \ref{sec_pile-up_pebbles}), runaway pile-ups of pebbles and silicate dust (Section \ref{sec_runaway_pile-up}), rock-to-ice ratio within the pile-up region (Section \ref{sec_rock-to-ice}), and implication for planet formation (Section \ref{sec_implication}).

\subsection{Comparison to previous works} \label{sec_comparison}
In most previous studies of the pile-up of silicate dust and pebbles around the snow line, $\alpha_{\rm acc}=\alpha_{\rm Dr}=\alpha_{\rm Dz}$ was assumed \citep{Dra17,Sch17}. Figure \ref{fig_map_variable} indicates that such parameters do not lead to a runaway pile-up of silicate dust inside the snow line, while a pile-up of pebbles outside the snow line could satisfy a condition for SI to take place for sufficiently large pebble mass flux of $F_{\rm p/g} > 0.8$ (Fig.~\ref{fig_map_variable}). Previous studies \citep{Dra17,Sch17} assumed $K=0$ and a runaway pile-up of pebbles was not reported (only a steady-state is reached with $\rho_{\rm p}/\rho_{\rm g} > 1$ for $F_{\rm p/g} > 0.8$), while it could take place when $K \neq 0$ with Drift-BKR \citep[see also][]{Hyo19}.

\subsection{Pile-up of silicate dust inside the snow line} \label{sec_pile-up_dust}
As discussed in Paper I, silicate dust can pile up just inside the snow line \citep[Fig.~\ref{fig_summary}; see also][]{Est16,Ida16,Sch17,Dra17,Hyo19,Gar20}. Our 1D simulations include the recycling process via sticking of the diffused dust onto icy pebbles beyond the snow line. It conserves the inward mass flux of silicate dust $F_{\rm d/g}$ across the snow line (all the diffused silicate dust eventually comes back to the snow line with pebbles), which indicates $F_{\rm d/g}=f_{\rm d/p} \times F_{\rm p/g}$ (here, $f_{\rm d/p}=0.5$ is the silicate fraction in the original pebbles at the outer boundary) is independent on $\alpha_{\rm acc}$ and $\alpha_{\rm Dr}$. The vertically integrated metallicity of silicate dust can be written as $Z_{\rm \Sigma_{\rm d}} \propto F_{\rm d/g} \times v_{\rm g}/v_{\rm d}$ and is independent on $\alpha_{\rm acc}$ as $v_{\rm d} \simeq v_{\rm g}$. The midplane solid-to-gas ratio is written as $\rho_{\rm d}/\rho_{\rm g} = Z_{\Sigma_{\rm d}} h_{\rm d/g}^{-1}$ ($h_{\rm d/g} \equiv H_{\rm d}/H_{\rm g}$). At the snow line, $h_{\rm d/g}$ is regulated by $\alpha_{\rm Dz}/\alpha_{\rm acc}$ (see Eq.~(\ref{eq_Hd_Ida1}) for $\alpha_{\rm Dz}=\alpha_{\rm Dr}$). Thus, both $\alpha_{\rm acc}=10^{-2}$ and $\alpha_{\rm acc}=10^{-3}$ cases have almost the same $\rho_{\rm d}/\rho_{\rm g}$ for the same $\alpha_{\rm Dz}/\alpha_{\rm acc}$ and $F_{\rm p/g}$ when $\rho_{\rm d}/\rho_{\rm g} \ll 1$ (the right numbers in the parenthesis in Fig.~\ref{fig_map_variable})\footnote{When $\rho_{\rm d}/\rho_{\rm g} > 1$, the back-reaction changes other quantities, such as the scale height of pebbles (Eq.~(\ref{eq_Hp})).}. 

Our 1D simulations (the right numbers in the parenthesis in Fig.~\ref{fig_map_variable}) and the results of an analytical formula that is calibrated by the Monte Carlo simulations (a color contour in Fig.~\ref{fig_map_variable} obtained from Paper I) show good consistency when pebble pile-up is not significant. Paper I neglected the recycling of water vapor onto pebbles, but it does not significantly affect the pile-up of silicate dust inside the snow line as the mass flux of silicate dust is a critical parameter for the pile-up.

\subsection{Pile-up of pebbles outside the snow line} \label{sec_pile-up_pebbles}

In this subsection, we discuss how pebbles pile up just outside the snow line and its dependence on $\alpha_{\rm acc}$. Outside the snow line, the pile-up of icy pebbles is locally enhanced due to two distinct recycling processes: the recondensation of water vapor and sticking of silicate dust \citep[Fig.~\ref{fig_summary}; see also][]{Ida16,Sch17,Dra17,Hyo19,Gar20}. The efficiency of the recycling process (i.e., $\alpha_{\rm Dr}$) is a critical parameter to regulate the pile-up of icy pebbles. Thus, unlike the case of silicate dust (Section \ref{sec_pile-up_dust}), the value $\rho_{\rm p}/\rho_{\rm g}$ is different for the same $\alpha_{\rm Dr}/\alpha_{\rm acc}(=\alpha_{\rm Dz}/\alpha_{\rm acc})$ and $F_{\rm p/g}$ at a different $\alpha_{\rm acc}$ (see the left and right panels in Fig.~\ref{fig_map_variable}).

The surface density of pebbles does not strongly depend on $\alpha_{\rm acc}$ as pebbles are only partially coupled to the gas flow \citep[$\tau_{\rm s,p} \sim 0.1$;][]{Oku12,Ida16} and the radial velocity of pebbles is mainly dominated by the gas drag (the first term in Eq.~(\ref{eq_vp})), while the surface density of the gas becomes smaller for larger $\alpha_{\rm acc}$ as $\Sigma_{\rm g} \propto 1/v_{\rm g} \propto 1/\alpha_{\rm acc}$. This indicates that $\rho_{\rm p}/\rho_{\rm g}$  is $\propto \alpha_{\rm acc}$ when the recycling processes are negligible.

As $F_{\rm p/g}$ increases or/and $\alpha_{\rm Dz}/\alpha_{\rm acc}$ decreases, the pile-up of pebbles tends to increase as the disk metallicity increases ($\Sigma_{\rm p} \propto F_{\rm p/g}$) and the scale height of pebbles decreases (Eqs.~(\ref{eq_Hp}) and (\ref{eq_Hd_Ida1})). However, when $\alpha_{\rm Dr}/\alpha_{\rm acc} \simeq 1$, an efficient recycling of water vapor and dust onto pebbles can change this dependence for some parameters (for example, the cases of $\alpha_{\rm acc}=\alpha_{\rm Dz}=10^{-3}$). Comparing the maximum $\rho_{\rm d}/\rho_{\rm g}$ and $\rho_{\rm p}/\rho_{\rm g}$, the pile-up of pebbles is more significant than that of silicate dust when $F_{\rm p/g}$ is sufficiently large and when the diffusion is an efficient process (i.e., $\alpha_{\rm Dr}/\alpha_{\rm acc} \sim 1$ cases) because the recycling of the vapor and the dust onto pebbles efficiently enhances the pile-up of pebbles \citep[][]{Hyo19}. The pile-up of silicate dust is more prominent when the diffusivity is weak (i.e., $\alpha_{\rm Dz}/\alpha_{\rm acc} \ll 1$ cases).

\subsection{Conditions for preferential pile-up of silicate dust versus pebbles} \label{sec_runaway_pile-up} 
Finally, we aim to understand under which conditions the runaway pile-ups of silicate dust or/and pebbles are favored around the snow line. 

The black lines in Fig.~\ref{fig_map_variable} indicate where pebbles cannot drift toward the snow line, corresponding to the ND region (see Section \ref{sec_ND}; $r=5$ au and $Ri=0.5$ are used here). In the case of $\alpha_{\rm acc}=10^{-2}$, the red-colored region is located above the black line, which implies that parameters in the $F_{\rm p/g}-\alpha_{\rm acc}-\alpha_{\rm Dz}$ space that would yield the runaway pile-up of silicate dust found in Paper~I in fact lie in the forbidden ND region. On the other hand, for $\alpha_{\rm acc}=10^{-3}$, the pile-up of silicate dust inside the snow line for $\alpha_{\rm Dz}(=\alpha_{\rm Dr}) < 10^{-5}$ and $0.1 < F_{\rm p/g} < 0.8$ found Paper I is outside of the ND region and confirmed by our simulations.

Generally, runaway pile-ups around the snow line occur over a broader range of parameters for $\alpha_{\rm acc}=10^{-3}$ than for $\alpha_{\rm acc}=10^{-2}$. Runaway pile-up of silicate dust is favored for a small $\alpha_{\rm Dz}/\alpha_{\rm acc}<10^{-3}$ (labeled by "Sil-RP"), while runaway pile-up of pebbles (labeled by "Peb-RP") only occurs for a limited range of parameters (large $F_{\rm p/g}$ and $\alpha_{\rm Dz}/\alpha_{\rm acc} \sim 1$). 

The results in Fig.~\ref{fig_map_variable} thus indicate that planetesimal formation around the snow line potentially occurs via a direct gravitational collapse of solids, to form icy or rocky planetesimal inside or outside the snow line, respectively, depending on the parameters of $F_{\rm p/g}$ and $\alpha_{\rm Dr}/\alpha_{\rm acc}$. However, the formation efficiencies of these two kinds of planetesimals strongly depend on the disk conditions. This emphasizes that detailed disk evolution models including a realistic treatment of pebble growth are required to understand planetesimal formation around the snow line.

\subsection{Rock-to-ice ratio within the pile-up region} \label{sec_rock-to-ice}
Around the snow line, the rock-to-ice mixing ratio can be modified due to sublimation, condensation, and the recycling effects \citep[][]{Hyo19}. Our numerical simulations show that pebbles just outside the snow line contain both ice and rock (the numbers below parenthesis in Fig.~\ref{fig_map_variable} indicate resulting rock fractions). The results indicate that $\alpha_{\rm Dr}/\alpha_{\rm acc}=10^{-1}$ cases tend to most efficiently increase the rock fraction from the original value (the 1:1 initial rock-ice ratio; Section \ref{sec_methods}) for a fixed $F_{\rm p/g}$. 

For a higher $\alpha_{\rm Dr}/\alpha_{\rm acc}$, the scale height of silicate dust becomes larger (Eq.~(\ref{eq_Hd_Ida1})), which leads to a less efficient sticking of silicate dust onto icy pebbles (its efficiency $\propto 1/H_{\rm d}$), leading to a lower rock fraction in pebbles. However, a higher $\alpha_{\rm Dr}/\alpha_{\rm acc}$ yields a more efficient outward diffusion of silicate dust released in the inner region, leading to a higher surface density of silicate dust outside the snow line, hence enhancing the silicate fraction in pebbles (the sticking efficiency $\propto \Sigma_{\rm d}$). 

Diff-BKR and Drift-BKR also play critical roles as these effects modify the diffusivity and radial motion of solids. As a result of a combination of these different effects, the numerical results show that the rock fraction in pebbles, generally increases from the initial value of $50$wt\% up to $\sim 80$wt\%, depending on $\alpha_{\rm Dr}/\alpha_{\rm acc}$ and $F_{\rm p/g}$ (Fig.~\ref{fig_map_variable}). When a runaway pile-up of pebbles occurs (labeled by Peb-RP in Fig.~\ref{fig_map_variable}), the rock fraction becomes $\sim 50$wt\%. This is because the runaway pile-up occurs by accreting pebbles that drift from the outer region. 

Thus, forming planetesimals from pebbles (by GI or SI) by runaway pile-up outside the snow line leads to a rock fraction being similar to that of the original pebbles formed in the outer region. The threshold $\rho_{\rm p}/\rho_{\rm g}$ for the occurrence of a streaming instability should be close to unity, within an order of magnitude \citep{Car15,Yan17}. As shown in Fig.~\ref{fig_map_variable}, lower values of this threshold could lead to the formation of rock-rich planetesimals even from the pile-up of pebbles outside of the ice line. This is for example the case for $\alpha_{\rm acc}=10^{-3}$ and $F_{\rm p/g}=0.3$. This may explain the high dust-to-ice ratio measured in comet 67P \citep{Oro20}. Planetesimals formed from the pile-up of silicate dust inside of the snow line should be purely rocky. 

\begin{figure*}[h]
	\centering
	\resizebox{\hsize}{!}{ \includegraphics{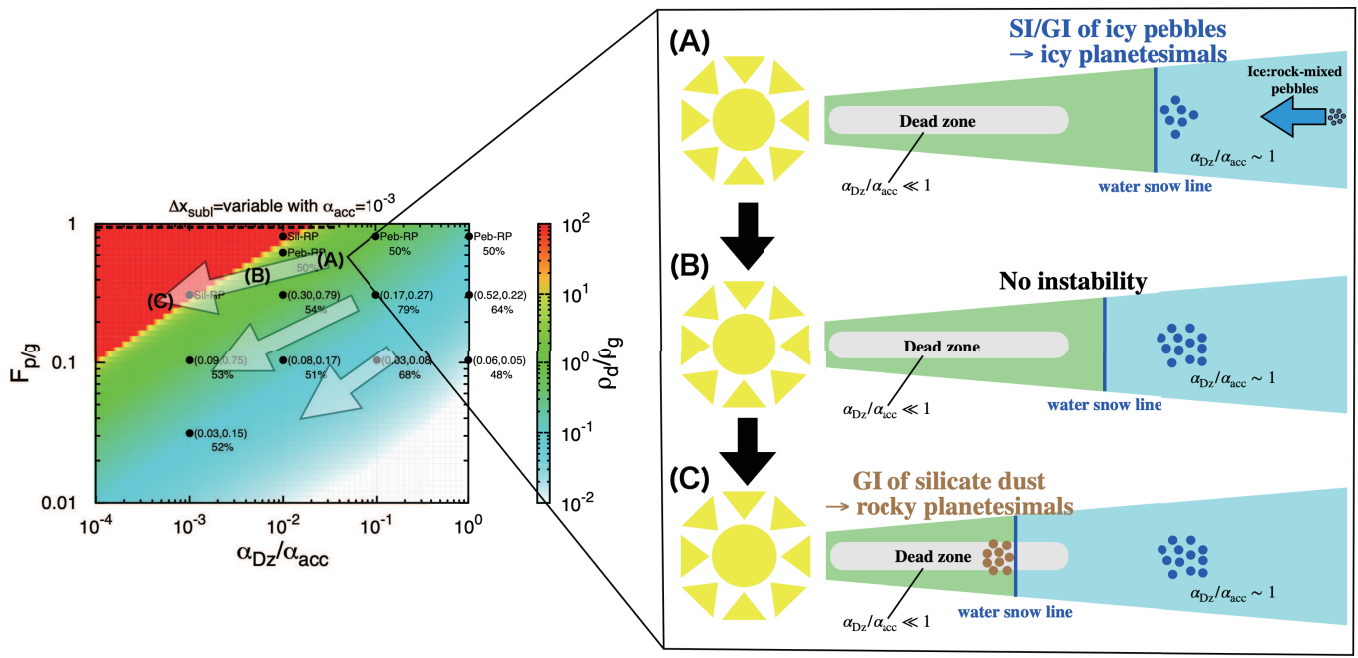} }
	\caption{Illustration showing a possible planet formation scenario. Here we envision a disk containing an MRI-inactive dead zone near the midplane ($\alpha_{\rm Dr}(=\alpha_{\rm Dz})/\alpha_{\rm acc} \ll 1$), while the outer disk is active ($\alpha_{\rm Dr}(=\alpha_{\rm Dz})/\alpha_{\rm acc} \sim1$). We also envision an initially high $F_{\rm p/g}$ value that becomes smaller with time, while the snow line moves inward. Our results of planetesimal formation around the snow line imply that the icy pebble pile-up (i.e., icy planetesimals formation by GI and/or SI; panel (A)) would be favored in the outer region ($\alpha_{\rm Dz}/\alpha_{\rm acc} \sim 1$) in the early stage of the evolution, while the formation of rocky planetesimals by GI of silicate dust would be favored in the later stage when the snow line has reached the dead zone ($\alpha_{\rm Dz}/\alpha_{\rm acc} \ll 1$; panel (C)). In the intermediate phase ($\alpha_{\rm Dz}/\alpha_{\rm acc} < 1$ but not $\alpha_{\rm Dz}/\alpha_{\rm acc} \ll 1$) and $F_{\rm p/g} < 0.6$, SI/GI of pebbles and GI of silicate dust would not be expected to occur (panel (B)). A diversity of evolutionary paths for protoplanetary disks would produce a diversity of planetary systems.}
\label{fig_implication}
\end{figure*}

\subsection{Implication for planet formation} \label{sec_implication}
The structure and evolution of protoplanetary disks remain poorly constrained. We propose with Fig.~\ref{fig_implication} an example scenario that would lead to the formation of both ice-rich and rock-rich planetesimals as seen in the solar system. 

The evolution of the disks is characterized by a decrease of the accretion rate, a decrease of viscous heating and therefore an inward migration of the snow line \citep{Oka11}. In the same time, $F_{\rm p/g}$ could decrease \citep{Ida16b}. The inner region of the disk may have an MRI-inactive "dead zone" near the disk midplane where turbulence and correspondingly diffusion are weak \cite[Fig.~\ref{fig_implication};][]{Gam96}. The radial boundary between the active zone in the outer disk region ($\alpha_{\rm Dr}(=\alpha_{\rm Dz})/\alpha_{\rm acc} \sim 1$) and the dead zone in the inner disk region ($\alpha_{\rm Dr}(=\alpha_{\rm Dz})/\alpha_{\rm acc} \ll1$) may smoothly change, that is, $\alpha_{\rm Dr}(=\alpha_{\rm Dz})/\alpha_{\rm acc}$ may gradually become smaller as the distance to the central star becomes smaller \citep[e.g.,][]{Bai16,Mor17}. 

Our results (see Fig.~\ref{fig_map_variable}) imply that in the early stage of the disk evolution when the snow line is located in the active outer region, a pile-up of icy pebbles leading to the formation of icy planetesimals by GI or SI may take place. As shown in Fig.~\ref{fig_implication}, late in the evolution, the migration of the snow line into the dead inner region would lead to the formation of rocky planetesimals. In-between, in an intermediate phase, the planetesimal formation around the snow line would be suppressed because of a combination of $F_{\rm p/g}$ and $\alpha_{\rm Dz}/\alpha_{\rm acc}$ that are insufficient to lead to a pile-up of solids in the disk midplane. 

As the observed structure and theoretically modeled evolutionary paths of protoplanetary disks are diverse, following a distinct evolutionary path in the $\alpha_{\rm acc}$--$\alpha_{\rm Dr}$--$\alpha_{\rm Dz}$--$F_{\rm p/g}$ space as well as that of the snow line could produce a diversity of the outcomes of the planetesimal formation around the snow line. A preferential formation of planetesimals only around the snow line would produce a localized narrow ring distribution of planetesimals.

\section{Summary} \label{sec_summary}

In this paper, to study the pile-up of solids around the snow line, we incorporated a realistic $H_{\rm d}$ (obtained from Paper I) and the KH-instability-considered $H_{\rm p}$ (Section \ref{sec_KH}), for the first time, into a 1D diffusion-advection code that includes the back-reactions to radial drift and diffusion of icy pebbles and silicate dust. The code takes into account ice sublimation, the release of silicate dust, and their recycling through the recondensation and sticking onto pebbles. We studied a much wider range of disk parameters (in the $F_{\rm p/g} - \alpha_{\rm acc} - \alpha_{\rm Dr} (=\alpha_{\rm Dz}$) space) than previous studies. Our main findings are as follows.

First, we derived the sublimation width of drifting icy pebbles, a critical parameter to regulate the scale height of silicate dust (Section \ref{sec_sub}). We also derived the scale height of pebbles, including the effects of KH instabilities (Section \ref{sec_KH}). 

Second, using analytical arguments, we identified a parameter regime in the $F_{\rm p/g}-\alpha_{\rm acc}-\alpha_{\rm Dz}$ space, in which pebbles cannot reach the snow line by stopping their radial drift due to Drift-BKR in a self-induced manner (the "no-drift" region; Section \ref{sec_ND}). 

Third, our 1D simulations showed that $\alpha_{\rm acc}=10^{-3}$ case is more favorable for the pile-up of solids around the snow line than that for $\alpha_{\rm acc}=10^{-2}$. The "no-drift" regime entirely covers parameter space of the runaway pile-up of silicate dust when $\alpha_{\rm acc}=10^{-2}$ case (left panel in Fig.~\ref{fig_map_variable}), preventing a runaway pile-up of silicate dust inside the snow line. In contrast, in the $\alpha_{\rm acc}=10^{-3}$ case, large $F_{\rm p/g}$ values are allowed and could lead to a runaway pile-up of silicate dust inside the snow line (right panel in Fig.~\ref{fig_map_variable}). In the case of $\alpha_{\rm acc}=10^{-3}$, both silicate dust and pebbles could experience runaway pile-ups inside and outside the snow line, forming rocky and icy planetesimals, respectively.

Forth, pebbles just outside the snow line is a rock-ice mixture (Section \ref{sec_rock-to-ice}). When a runaway pile-up of pebbles occurs just outside the snow line, the local rock-to-ice ratio becomes similar to that of pebbles formed at the outer region -- i.e., the resultant planetesimal composition would be water-bearing materials, if planetesimals ultimately formed from such pebbles. In contrast, almost pure rocky planetesimals formed from silicate dust, if GI occurs inside the snow line. 

Lastly, we discussed the implication of the runaway pile-ups around the snow line to planet formation (Section \ref{sec_implication}). A diversity of outcomes in terms of planetesimal formation around the snow line would occur for a diversity of protoplanetary disks. Detailed prescriptions of disk evolution and pebble growth are necessary.

\begin{acknowledgements}
We thank Dr. Chao-Chin Yang for discussions. We thank Dr. Beibei Liu for his constructive comments that improved the manuscript. RH was supported by JSPS Kakenhi JP17J01269 and 18K13600. RH also acknowledges JAXA's International Top Young program. TG was partially supported by a JSPS Long Term Fellowship at the University of Tokyo. SI was supported by MEXT Kakenhi 18H05438. SO was supported by JSPS Kakenhi 19K03926 and 20H01948. ANY was supported by NASA Astrophysics Theory Grant NNX17AK59G and NSF grant AST-1616929.
\end{acknowledgements}

\bibliography{planetesimals}

\appendix

\section{Dependence of dust pile-up upon $H_{\rm d}$} \label{sec_app_Hd}

Silicate dust is released during the sublimation of the drifting icy pebbles approaching the snow line. Different works have adopted different models of the scale height of silicate dust $H_{\rm d}(r)$ in a 1D-radial accretion disk \citep[e.g.,][]{Ida16,Sch17,Hyo19}. The midplane density of silicate dust inside the snow line is calculated by $\rho_{\rm d} = \Sigma_{\rm d}/\sqrt{2\pi}H_{\rm d}$, which is also used to describe the strength of the back-reaction of the gas onto their motion (e.g., Eq.~(\ref{eq_vd})). Therefore, the description of $H_{\rm d}$ is critical for the back-reaction and to evaluate the pile-up of dust in the midplane. In this section, we demonstrate how different models of $H_{\rm d}$ affect the dust pile-up inside the snow line.

 \cite{Ida16} neglected the effects of the vertical stirring of the released dust and assumed that the dust keeps the same scale height as that of pebbles, that is, $H_{\rm d}(r) = H_{\rm p}(r_{\rm snow})$, where $H_{\rm p}(r_{\rm snow})$ is the scale height of pebbles at the snow line. In contrast, \cite{Sch17} assumed that the released silicate dust is instantaneously stirred to the vertical direction and has the same scale height as that of the gas, that is, $H_{\rm d}(r) = H_{\rm g}(r)$.

Because silicate dust is released from sublimating pebbles, \cite{Hyo19} modeled that the released silicate dust has the same scale height as that of pebbles at the snow line and is gradually stirred to the vertical direction as a function to the distance to the snow line as
\begin{equation}
\label{eq_silicate_H}
	H_{\rm d} = \left( 1+ \frac{ \tau_{\rm s,p,snow} }{\alpha_{\rm Dz}} \times e^{ \frac{-\Delta t}{t_{\rm mix}}} \right)^{-1/2} H_{\rm g} 
\end{equation}
where a notation of ''snow'' indicates values at the snow line. $\Delta t=|(r-r_{\rm snow})/v_{\rm d}|$. $t_{\rm mix}=(H_{\rm g}/l_{\rm mfp})^2/\Omega_{\rm K}$ is the diffusion/mixing timescale to the vertical direction at the snow line and $ l_{\rm mfp}=\sqrt{\alpha_{\rm Dz}}H_{\rm g}$ is the mean free path of turbulent blobs. $\Delta t$ and $t_{\rm mix}$ are calculated by using the physical values at the snow line and we use $\tau_{\rm s,p,snow}=0.1$.

\begin{figure*}[t]
	\centering
	\resizebox{\textwidth}{!}{ \includegraphics{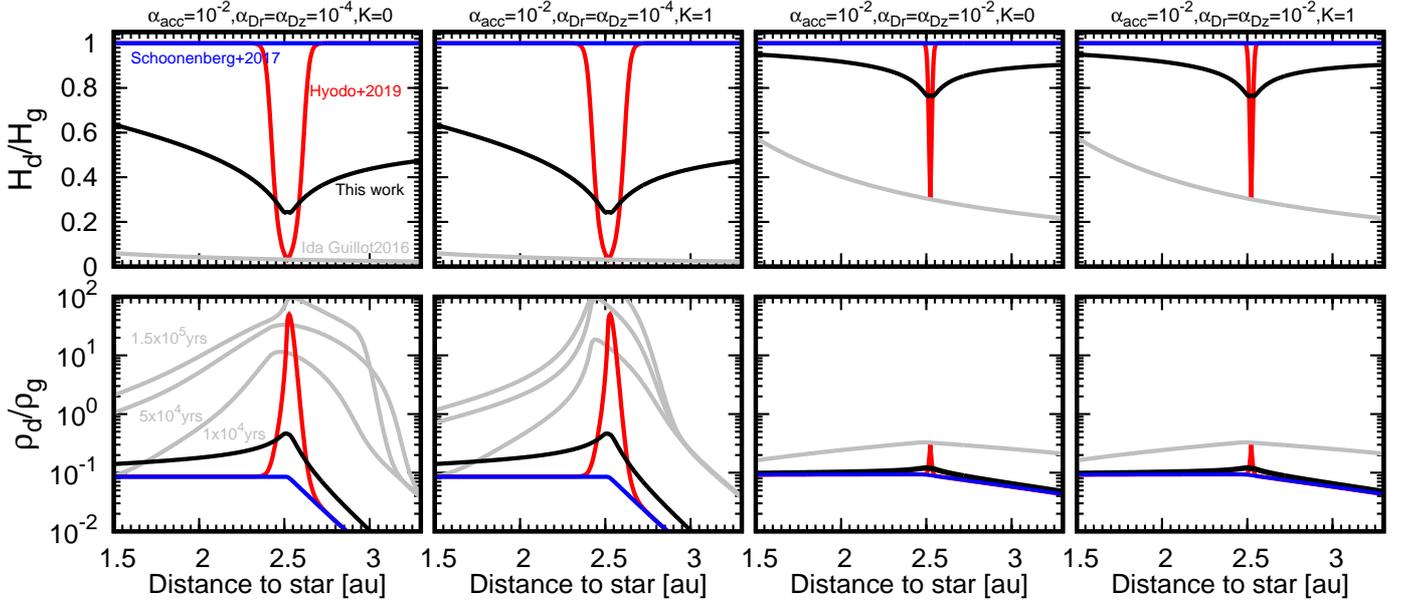}}
	\caption{Scale height of silicate dust in a unit of that of the gas (top panels) and midplane dust-to-gas ratio (bottom panels) for different models of $H_{\rm d}$ (Case of $F_{\rm p/g}=0.3$ and $\tau_{\rm s,p}=0.1$ at 3 au). Here, Drift-BKR and Diff-BKR for pebbles are neglected, whereas Drift-BKR and Diff-BKR ($K=0$ or $1$) for silicate dust are included. Left two panels are the cases of $\alpha_{\rm acc}=10^{-2}$ and $\alpha_{\rm Dr}=\alpha_{\rm Dz}=10^{-4}$ with $K=0$ or $1$, respectively. Right two panels are the cases of $\alpha_{\rm acc}=10^{-2}$ and $\alpha_{\rm Dr}=\alpha_{\rm Dz}=10^{-2}$ with $K=0$ or $1$, respectively. The gray lines are the case of \cite{Ida16}. The blue lines are the case of \cite{Sch17}. The red lines are the case of \cite{Hyo19} and the results show runaway pile-up when $\alpha_{\rm acc}=10^{-2}$ and $\alpha_{\rm Dr}=\alpha_{\rm Dz}=10^{-4}$ with $K=1$. The black lines are the case of this work (Eq.~(\ref{eq_Hd_Ida1})). In the cases of the gray lines \citep[i.e.,][]{Ida16}, pile-ups occur in a runaway fashion for the left two panels, and the time-evolutions are shown for $t=1 \times 10^{4}$ years, $t=5 \times 10^{4}$ years, and $t=1.5 \times 10^{5}$ years. The other cases reach a steady-state within $t=5 \times 10^{5}$ years. Here, the recycling of water vapor and silicate dust onto pebbles are not included to focus on the different models of $H_{\rm d}$.}
	\label{fig_Hsil}
\end{figure*}

Fig.~\ref{fig_Hsil} shows different models of the scale height of silicate dust and their resultant pile-ups of silicate dust, including a more realistic model used in this work (Equation (\ref{eq_Hd_Ida1}) which is originally derived in Paper I).

The model of \cite{Sch17}, $H_{\rm d}=H_{\rm g}$, underestimated pile-up of silicate dust when $\alpha_{\rm Dz} \ll \alpha_{\rm acc}$, while it correctly evaluated when  $\alpha_{\rm Dz} \simeq \alpha_{\rm acc}$. \cite{Ida16}, $H_{\rm d}=H_{\rm p,snow}$, overestimated the pile-up of silicate dust regardless of the value of $\alpha_{\rm Dz}$. \cite{Hyo19} overestimated pile-up of silicate dust near the snow line. 

\section{Particle size and Stokes number} \label{sec_appendix_size}

\begin{figure}[h]
	\centering
	\resizebox{\hsize}{!}{ \includegraphics{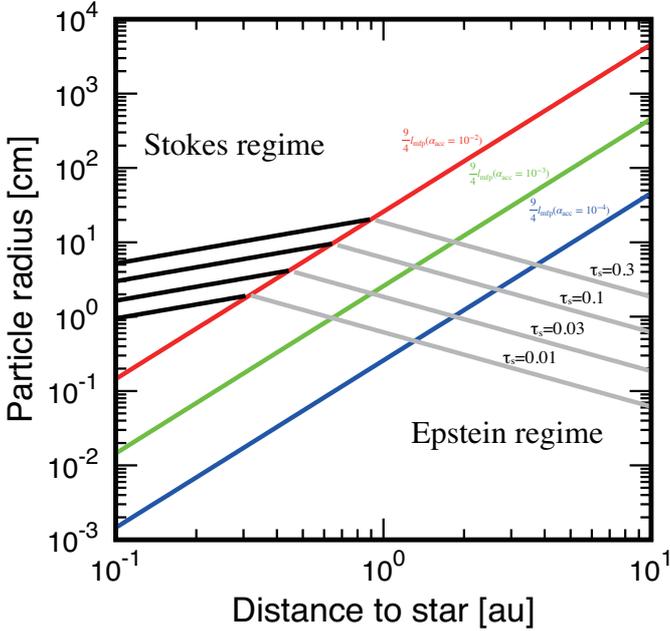} }
	\caption{Particle size as a function of distance to the star. The red, green, and blue lines represent $r_{\rm p}=\frac{9}{4}l_{\rm mfp}$ for $\alpha_{\rm acc}=10^{-2}$, $10^{-3}$, and $10^{-4}$, respectively (Eq.~(\ref{eq_mfp})). The Stokes regime is where $r_{\rm p}> \frac{9}{4}l_{\rm mfp}$, while the Epstein regime is where $r_{\rm p} < \frac{9}{4}l_{\rm mfp}$. Four black and gray lines represent particle size whose $\tau_{\rm s}=0.3$, $0.1$, $0.03$, and $0.01$ in the Stokes regime and the Epstein regime with $\alpha_{\rm acc}=10^{-2}$, respectively. Here, $\dot{M}_{\rm g}=10^{-8} M_{\odot}$/year and $T(r)=150\left( r/3{\rm au} \right)^{-1/2}$ are used.} 
	\label{fig_epstein_stokes}
\end{figure}

The Stokes number of a particle $\tau_{\rm s}(r)$ depends on the local gas structure and $\tau_{\rm s}(r)$ is written by using the stopping time $t_{\rm s}$ as $\tau_{\rm s}(r)=t_{\rm s}(r)\Omega_{\rm K}(r)$. The stopping time represents the relaxation timescale of particle momentum through the gas drag. Here, we consider two regimes of the stopping time and $\tau_{\rm s}(r)$ is given as 
\begin{align}
\label{eq_epstein}
	\tau_{\rm s}(r) & = t_{\rm s}(r)\Omega_{\rm K}(r) = \frac{\rho_{\rm p}r_{\rm p}}{\rho_{\rm g}v_{\rm th}} \Omega_{\rm K} \hspace{1em} {\rm Epstein:} r_{\rm p} < \frac{9}{4} l_{\rm mfp} \\
\label{eq_stokes}
	\tau_{\rm s}(r) & = t_{\rm s}(r)\Omega_{\rm K}(r) = \frac{4\rho_{\rm p}r_{\rm p}^2}{9\rho_{\rm g}v_{\rm th} l_{\rm mfp}} \Omega_{\rm K} \hspace{1em} {\rm Stokes:} r_{\rm p} > \frac{9}{4} l_{\rm mfp} 
\end{align}
where $v_{\rm th}(r)=\sqrt{ 8/\pi} c_{\rm s}(r)$ is the thermal velocity, $r_{\rm p}$ is the particle radius, and $\rho_{\rm p}=1.5$ g cm$^{-3}$ is the particle internal density, and $\rho_{\rm g}=\Sigma_{\rm g}/\sqrt{2\pi}H_{\rm g}$ is the gas spatial density, respectively. The mean free path of the gas $l_{\rm mfp}(r)$ is given as 
\begin{equation}
\label{eq_mfp}
	l_{\rm mfp}(r) = \frac{\mu_{\rm g} m_{\rm proton}}{\sqrt{2}\rho_{\rm g}(r)\sigma_{\rm mol}}
\end{equation}
where $\sigma_{\rm mol}=2.0\times10^{-15}$ cm$^{2}$ is the collisional cross section of the gas molecules. We note that $l_{\rm mfp}$ depends on the local gas structure and is a function of $\alpha_{\rm acc}$. 

Rewriting Eqs.~(\ref{eq_epstein}) and (\ref{eq_stokes}) give particle sizes $r_{\rm p}$ in the Epstein and Stokes regimes for a fixed $\tau_{\rm s}$. Fig.~\ref{fig_epstein_stokes} shows critical particle sizes of $r_{\rm p}=\frac{9}{4}l_{\rm mfp}$ for different $\alpha_{\rm acc}$ and particle sizes in either Epstein and Stokes regimes for a fixed $\tau_{\rm s}$. Because $\nu_{\rm acc} \propto \alpha_{\rm acc}r^{\frac{3}{2}-\beta}$, $H_{\rm g} \propto r^{\frac{3-\beta}{2}}$, $\Sigma_{\rm g} \propto \alpha_{\rm acc}^{-1} r^{\beta-\frac{3}{2}}$, $\rho_{\rm g} \propto \alpha_{\rm acc}^{-1} r^{\frac{3}{2}\beta-{3}}$, and $v_{\rm th} \propto r^{-\frac{\beta}{2}}$, the particle size in the Epstein regime becomes $r_{\rm p} \propto \alpha_{\rm acc}^{-1} r^{\beta-\frac{3}{2}}$, that is, $r_{\rm p} \propto r^{-1}$ for $\beta=1/2$ (the gray lines in Fig.~\ref{fig_epstein_stokes}). Because $l_{\rm mfp} \propto 1/\rho_{\rm g} \propto \alpha_{\rm acc} r^{3-\frac{3}{2}\beta}$, the particle size in the Stokes regime becomes $r_{\rm p} \propto r^{\frac{3}{4}-\frac{\beta}{4}}$, which is independent on $\alpha_{\rm acc}$, that is, $r_{\rm p} \propto r^{\frac{5}{8}}$ for $\beta=1/2$ (the black lines in Fig.~\ref{fig_epstein_stokes}).

\section{Dependence on the sublimation width} \label{sec_dep_subl}
\begin{figure*}[t]
	\centering
	\resizebox{\hsize}{!}{ \includegraphics{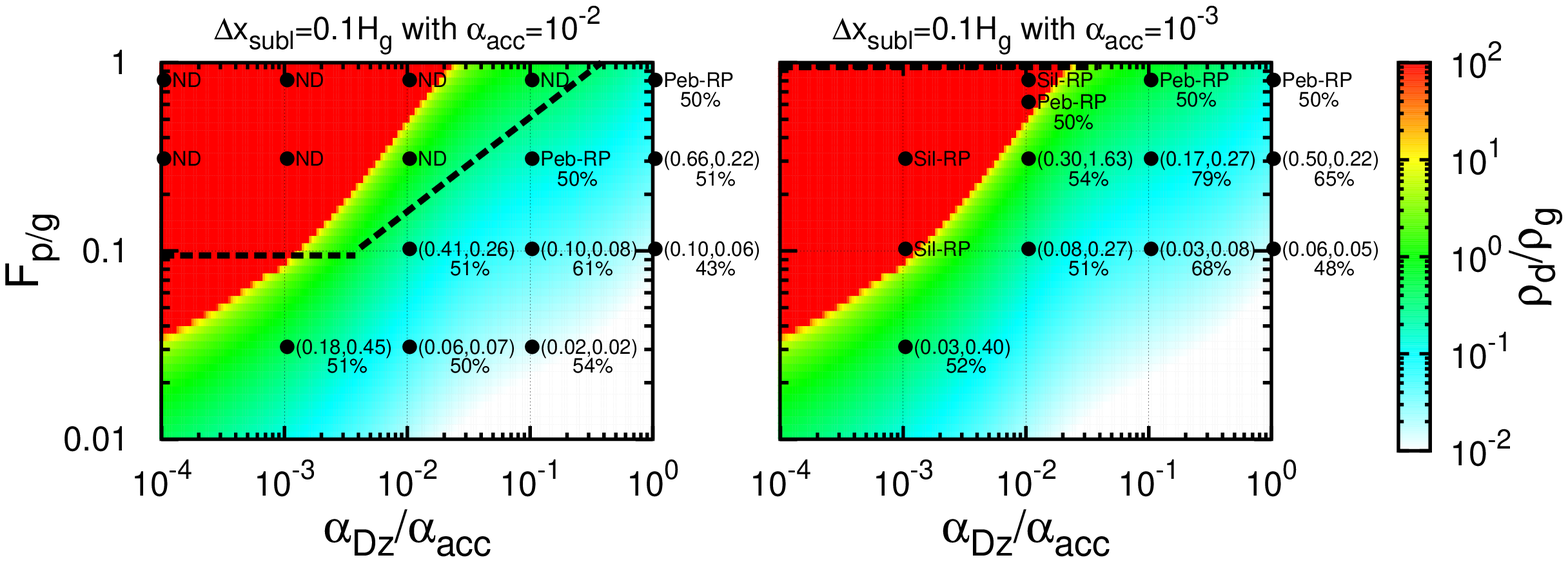} }
	\caption{Same as Fig.~\ref{fig_map_variable}, but for the case of a constant $\Delta x_{\rm subl}=0.1H_{\rm g}$.}
	\label{fig_map_fixed}
\end{figure*}

Here, we discuss the dependence on the sublimation width $\Delta x_{\rm subl}$. The detailed derivation of the sublimation width is discussed in Section \ref{sec_sub}. In Fig.~\ref{fig_map_variable}, a realistic variable sublimation width is used (Eq.~(\ref{eq_fitting})). In contrast, Fig.~\ref{fig_map_fixed} is the same as Fig.~\ref{fig_map_variable} but the sublimation width is fixed ($\Delta x_{\rm subl}=0.1H_{\rm g}$).

As the sublimation width $\Delta x_{\rm subl}$ becomes larger, the average scale height of silicate dust becomes larger because the silicate dust released in a wider radial width diffuses vertically and radially mixes (Eq.~(\ref{eq_Hd_Ida1})), leading to a lower concentration of silicate dust in the disk midplane. This effect is more significant for small $\alpha_{\rm Dr}(=\alpha_{\rm Dz})/\alpha_{\rm acc}$ (i.e., the advection-dominated regime) as both radial and vertical mixings are ineffective, while large $\alpha_{\rm Dr}(=\alpha_{\rm Dz})/\alpha_{\rm acc}$ (i.e., the diffusion-dominated regime) case is rarely affected because the mixing effects "erase" the information of the initial distribution of silicate dust. Thus, the parameter space for runaway pile-up of silicate dust is reduced for small $\alpha_{\rm Dr}(=\alpha_{\rm Dz})/\alpha_{\rm acc}$ in the case of a variable $\Delta x_{\rm sub}$ (Eq.~(\ref{eq_fitting})) compared to the case of the fixed $\Delta_{\rm x}=0.1H_{\rm g}$ (compare Fig.~\ref{fig_map_fixed} to Fig.~\ref{fig_map_variable}).

\end{document}